\RequirePackage{silence}
\documentclass[fleqn,usenatbib,useAMS]{mnras} 
\usepackage{graphicx}
\usepackage{amsmath}

\usepackage{amsfonts}
\usepackage{float}
\usepackage{bm}
\usepackage[perpage,symbol*]{footmisc}
\usepackage{ae,aecompl}
\usepackage{array}
\usepackage{soul}
\usepackage{mathtools}
\usepackage{multirow}
\usepackage[utf8]{inputenc}
\usepackage{booktabs}
\usepackage{graphicx}

\DeclareRobustCommand{\appropto}{\mathrel{\vcenter{
		\offinterlineskip\halign{\hfil$##$\cr 
			\propto\cr\noalign{\kern2pt}\sim\cr\noalign{\kern-2pt}}}}}

\title[El Gordo $-$ a massive blow to $\Lambda$CDM cosmology]{A massive blow for $\Lambda$CDM $-$ the high redshift, mass, and collision velocity of the interacting galaxy cluster El Gordo contradicts concordance cosmology} 

\author[E.~Asencio, I.~Banik and P.~Kroupa]{Elena Asencio$^{1}$\thanks{Email: \href{mailto:s6elena@uni-bonn.de}{s6elena@uni-bonn.de} (Elena Asencio) \newline $~~~~~~~~~$ \href{mailto:ibanik@astro.uni-bonn.de}{ibanik@astro.uni-bonn.de} (Indranil Banik)}, Indranil Banik$^{1}$, and Pavel Kroupa$^{1,2}$  \vspace{10pt} \\
$^{1}$Helmholtz-Institut f\"ur Strahlen und Kernphysik (HISKP), University of Bonn, Nussallee 14$-$16, D-53115 Bonn, Germany \\
$^{2}$Astronomical Institute, Faculty of Mathematics and Physics, Charles University, V Hole\v{s}ovi\v{c}k\'ach 2, CZ-180 00 Praha 8, Czech Republic}

\pubyear{2020}
\begin{document}
\label{firstpage}
\pagerange{\pageref{firstpage}--\pageref{lastpage}}

\maketitle

\begin{abstract} 

El Gordo (ACT-CL J0102-4915) is an extremely massive galaxy cluster ($M_{200} \approx 3 \times 10^{15}$ $M_{\odot}$) at redshift $z = 0.87 $ composed of two subclusters with mass ratio 3.6 merging at speed $V_{infall} \approx 2500$~km/s. Such a fast collision between individually rare massive clusters is unexpected in Lambda cold dark matter ($\Lambda$CDM) cosmology at such high $z$. However, this is required for non-cosmological hydrodynamical simulations of the merger to match its observed properties (Zhang et al. 2015). Here, we determine the probability of finding a similar object in a $\Lambda$CDM context using the Jubilee simulation box with side length $6 \, h^{-1}$~Gpc. We search for galaxy cluster pairs that have turned around from the cosmic expansion with properties similar to El Gordo in terms of total mass, mass ratio, redshift, and collision velocity relative to virial velocity. We fit the distribution of pair total mass quite accurately, with the fits used in two methods to infer the probability of observing El Gordo in the surveyed region. The more conservative (and detailed) method involves considering the expected distribution of pairwise mass and redshift for analogue pairs with similar dimensionless parameters to El Gordo in the past lightcone of a $z = 0$ observer. Detecting one pair with its mass and redshift rules out $\Lambda$CDM cosmology at $6.16\sigma$. We also use the results of Kraljic \& Sarkar (2015) to show that the Bullet Cluster is in $2.78\sigma$ tension once the sky coverage of its discovery survey is accounted for. Using a $\chi^2$ approach, the combined tension can be estimated as $6.43 \sigma$. Both collisions arise naturally in a Milgromian dynamics (MOND) cosmology with light sterile neutrinos.

\end{abstract}


\section{Introduction}
\label{Introduction}

According to the hierarchical structure formation paradigm, smaller structures formed first and then merged into larger structures, eventually forming galaxy clusters \citep{Davis_1985}. These are the largest gravitationally bound structures in the Universe. The efficiency and timescale proposed for their formation varies depending on the cosmological model adopted. The most generally accepted model nowadays is Lambda-Cold Dark Matter \citep[$\Lambda$CDM;][]{Ostriker_1995}. $\Lambda$CDM is built on the assumption of a homogeneous and isotropically expanding Universe on very large scales, with all its matter-energy content appearing at very early times in a hot Big Bang. Two other elements had to be added in order to explain certain astronomical phenomena: the unexpected faintness of distant Type Ia supernovae at fixed redshift $z$ suggested that the Universe was undergoing an accelerated expansion, which implied that Einstein's field equations should have a non-zero cosmological constant $\Lambda$ \citep{Riess_1998}; and the observed flat rotation curves of galaxies \citep[e.g.][]{Babcock_1939, Rubin_1970, Rogstad_1972, Bosma_1978} indicated that the Newtonian gravity of the baryons alone was not enough to hold them together, which led to the postulation of undetected mass that would surround galaxies in the form of a halo \citep{Ostriker_Peebles_1973}. Gravitational microlensing results show that the Galactic halo cannot consist of compact objects with planetary or stellar mass \citep[e.g.][]{MACHO_2000, EROS_2007}, so it is instead thought to consist of non-baryonic particles that do not exist in the well-tested standard model of particle physics. Null detection of non-gravitational signals from the postulated CDM particles places stringent constraints on their allowed properties \citep[e.g.][]{Abazajian_2020, Hoof_2020}. Nonetheless, $\Lambda$CDM has proven successful in explaining e.g. the cosmic microwave background \citep[CMB;][]{Planck_XXVII_2013}.

However, its predicted efficiency for the formation of galaxy clusters has been challenged by observations of several massive galaxy clusters at high $z$ \citep[e.g.][]{Mullis_2005, Ebeling_2007, Lamer_2008, Jee_2009, Menanteau_2010, Foley_2011, Stanford_2012, Stalder_2013, Buddendiek_2015, Tozzi_2015, Wang_2016, Miller_2018, Wen_2018}. The most massive cluster at $z > 1$ is SPT-CL J2106-5844, with $z = 1.132^{+0.002}_{-0.003}$ and mass $M_{200} = \left( 1.27 \pm 0.21 \right) \times 10^{15} M_\odot$ \citep{Foley_2011}. This was confirmed by the more recent study of \citet{Kim_2019}, which gives a slightly lower mass estimate of  $M_{200} = \left( 1.04 ^{+0.33}_{-0.30} \right) \times 10^{15} M_\odot$. Their figure~6 shows two weak lensing peaks, indicating the system is likely a merger product. Further studies are required to pin down the pre-merger configuration. The Bullet Cluster (1E 0657-56) at redshift $z = 0.296$ and El Gordo (ACT-CL J0102-4915) at $z = 0.870$ are the galaxy cluster encounters most well known for having a very high relative velocity \citep{Markevitch_2004} and for being the most massive distant galaxy cluster \citep{Menanteau_2012}, respectively. These observations suggest that the formation of large structures took place earlier than expected in $\Lambda$CDM.

On even larger scales of tens of Mpc, galaxy clusters are often part of a supercluster. The existence of very massive superclusters at high redshift also seems to pose problems for the $\Lambda$CDM model \citep{Cucciati_2018, Hayashi_2019}. The former work identified a $4.8 \times 10^{15} M_\odot$ supercluster at $z = 2.45$, which is consistent with the predicted mass function at $z = 1$. Though a relevant simulation snapshot was unavailable for a more meaningful comparison at $z = 2.5$, the rapidly rising number density of such massive superclusters around the epoch $z = 1$ \citep[figure~2 of][]{Lim_2014} suggests that there would be significant tension with the observations of \citet{Cucciati_2018}.

If structure formation is enhanced compared to $\Lambda$CDM expectations, then not only overdensities but also underdensities should be more pronounced than expected. There is actually strong evidence for a large local underdensity extending out to $\approx$~300 Mpc \citep{KBC_2013}. This is in $6.04\sigma$ tension with $\Lambda$CDM expectations \citep[section~2 of][]{Haslbauer_2020}. Those authors showed that such a large and deep void would cause the local expansion rate of the Universe to exceed the average by $\approx 11\%$ (see their equation 5). This would nicely resolve the Hubble tension, i.e. the difficulty in reconciling the observed CMB anisotropies \citep{Aiola_2020, Planck_VI_2020} with multiple independent determinations of the local expansion rate \citep[][and references therein]{Riess_2020}.

The problem of early structure formation also extends down to galaxy scales \citep[e.g.][]{Kang_2015, Girelli_2019, Forrest_2020} and their central super massive black holes (SMBHs). In this respect, the recent discovery of the luminous quasar J1007+2115 at $z = 7.515$ is particularly noteworthy $-$ it contains an SMBH of mass $\left( 1.5 \pm 0.2 \right) \times 10^9 M_{\odot}$. This discovery challenges models of SMBH growth, since these would require a seed black hole of $\geq 10^4 M_{\odot}$ at $z = 30$ to explain the observations \citep{Yang_2020}.

Many of the publications addressing this rapid structure formation tension on the cluster scale focus on the Bullet Cluster \citep{Tucker_1995} as one of the main objects with which to test the likelihood of finding such massive high-$z$ clusters in a $\Lambda$CDM context \citep{Hayashi_2006, Angus_2008, Lee_2010, Thompson_Nagamine_2012, Katz_2013, Lage_2014, Watson_2014, Kraljic_2015}. This is because the Bullet Cluster consists of two colliding galaxy clusters with a collisional trajectory roughly normal to the line of sight, making it easier to study the interaction. It also has a high mass for a cluster at $z = 0.296~-$ the main cluster has $M_{200} \approx 1.9 \times 10^{15}$ $M_{\odot}$, while the subcluster mass is $M_{200} \approx 2.6 \times 10^{14}$ $M_{\odot}$ \citep{Lage_2014}. Most importantly, its shock and subcluster velocity were initially calculated to be $v \approx 4500$~km/s \citep{Markevitch_2004}, making this cluster quite problematic for $\Lambda$CDM \citep[e.g.][]{Thompson_Nagamine_2012}. More recent hydrodynamical simulations show that a lower infall velocity of $v \approx 3000$~km/s with an impact parameter of $b = 0.256$~Mpc is also compatible with observations \citep{Lage_2014}. This alleviated the tension significantly, since the Bullet Cluster infall velocity is now below its escape velocity (see their table~1). Nevertheless, according to \citet{Kraljic_2015}, this cluster is still considered a rare object for a $\Lambda$CDM cosmology: only 0.1 similar systems can be expected within $z < 0.3$ if we require that the collision has already occurred, as noticed by their referee. The Bullet Cluster was discovered in a survey covering only 5.4\% of the whole sky, making it a $2.78\sigma$ outlier for $\Lambda$CDM (Section~\ref{Combined_with_BC}). Finding yet more objects like this could imply a problem for the model.

The second aforementioned cluster of interest, El Gordo, presents significantly more striking features than the Bullet Cluster. El Gordo was observed for the first time by \citet{Menanteau_2010} in the Atacama Cosmology Telescope's 2010 survey of galaxy clusters, which were selected via the Sunyaev-Zel'dovich (SZ) effect \citep{Sunyaev_1970}. El Gordo consists of two merging subclusters at $z = 0.87$ with cores separated by a projected distance of $\approx 700$~kpc \citep[figure 5 of][]{Jee_2014}, a high infall velocity $V_{inf} \approx 2500$~km/s, and total mass $M_{200} \approx \left( 2-3 \right) \times 10^{15}$ $M_{\odot}$. These values are estimated using both observations \citep{Menanteau_2012, Zitrin_2013, Jee_2014} and idealized hydrodynamical simulations of the interaction \citep{Molnar_2015, Zhang_2015}. This makes El Gordo the most massive galaxy cluster at $z > 0.6$. Due to the highly energetic interaction of its two subclusters, it is also the hottest, most X-ray luminous, and brightest SZ effect galaxy cluster at this redshift \citep{Menanteau_2012}. Its bright X-ray emission has a single peak and a characteristic elongated shape with two faint tails that extend beyond this peak. The X-ray emission can be detected even in the outer regions of the merging cluster. Other distinctive features of El Gordo are the large offsets between the X-ray and SZ centroids and between the SZ centroid and the mass centre of the main cluster. These offsets are expected features in high velocity merging galaxy clusters \citep{Molnar_2012}, supporting the published high infall velocity estimates for El Gordo.

Even though the mass and redshift might appear as the most obvious oddities of El Gordo, we show in Section~\ref{Infall_velocity_condition} that its infall velocity significantly exceeds the escape velocity according to the \citet{Zhang_2015} results. This is likely not the case in the Bullet Cluster \citep[table~1 of][]{Lage_2014}. For two objects to be infalling faster than their escape velocity, there must have been a third object that pushed them towards each other $-$ any peculiar velocities at high $z$ are rapidly redshifted away by Hubble drag. If finding two objects this massive sufficiently close to each other is already rare, finding three objects like this is extremely unlikely. Moreover, the mass ratio between the two components of El Gordo is almost certainly in the range $2-5$ \citep{Zhang_2015}, with their best guess being 3.6 while other studies suggest $\approx 2$ \citep{Jee_2014, Molnar_2015}. This makes El Gordo a major merger. Such events are expected to be much less common than minor mergers like the Bullet Cluster, where the mass ratio is $\approx 7$ \citep{Lage_2014}. To summarize, El Gordo is a much more exceptional object than the Bullet Cluster in terms of the total mass, mass ratio, redshift, and infall velocity.

Deeper surveys are needed to find higher redshift objects, which usually requires a trade-off against the surveyed area. El Gordo was found in a very small sky region of 455~$deg^2$ \citep[1.1\% of the whole sky;][]{Menanteau_2010}. However, the survey that obtained the first constraints on El Gordo's properties was slightly larger: it covered 755~$deg^2$ \citep[1.8\% of the sky;][]{Menanteau_2012}. In our analysis, we adopt the latter value to be more conservative.

Despite its extreme nature, there are not so many works analysing the chance of finding El Gordo in a $\Lambda$CDM universe. This is partly because it was discovered quite recently, at least in comparison to the Bullet Cluster. Due to its large mass, a very large simulation box is needed to study El Gordo in a statistically meaningful way. The few attempts to do so \citep{Menanteau_2012, Katz_2013, Jee_2014, Sahlen_2016} do not provide a very detailed analysis. Except for \citet{Katz_2013}, these works consider El Gordo to be a single massive object instead of two massive objects at a very close distance infalling at high speed. The latter configuration is significantly less likely for clusters which are individually rather rare. Using the exclusion curve method of \citet{Mortonson_2011}, \citet{Menanteau_2012} and \citet{Jee_2014} found that El Gordo is a rare object but is not in tension with $\Lambda$CDM. \citet{Jee_2014} also added that, since El Gordo is only marginally allowed in $\Lambda$CDM, a more accurate measurement of its properties could give rise to significant tension. \citet{Sahlen_2016} estimated the likelihood of finding an El Gordo analogue according to the cluster and void mass functions predicted by $\Lambda$CDM. Using a Markov chain Monte Carlo method, they also concluded that El Gordo is not in tension with $\Lambda$CDM. \citet{Katz_2013} commented on the low likelihood of this object arising in a $\Lambda$CDM universe according to the mass-redshift relation of \citet{Mortonson_2011}. Importantly, \citet{Katz_2013} mentioned that the probability is even lower if one takes into account the high infall velocity of El Gordo, which can make it a more serious problem for $\Lambda$CDM. Indeed, in their cubic simulation box with sides of $1 \, h^{-1}$ co-moving Gpc (cGpc), they found no pairs with the required mass, redshift, and collision velocity.

Finding analogues to massive objects like El Gordo in a cosmological simulation requires a very large box size to reliably estimate the occurrence rate. This precludes the use of hydrodynamical simulations. But with only dark matter, it is not possible to try and match e.g. the observed X-ray morphology of El Gordo. Instead, we rely on idealized hydrodynamical simulations to tell us the pre-merger configuration. We then seek pairs in the cosmological simulation with similar total mass, redshift, and collision velocity to the El Gordo progenitors. These properties should be little affected by small-scale baryonic processes, and so should be well reproduced in a dark matter-only simulation.

\begin{figure}
	\centering
	\includegraphics[width = 8.5cm] {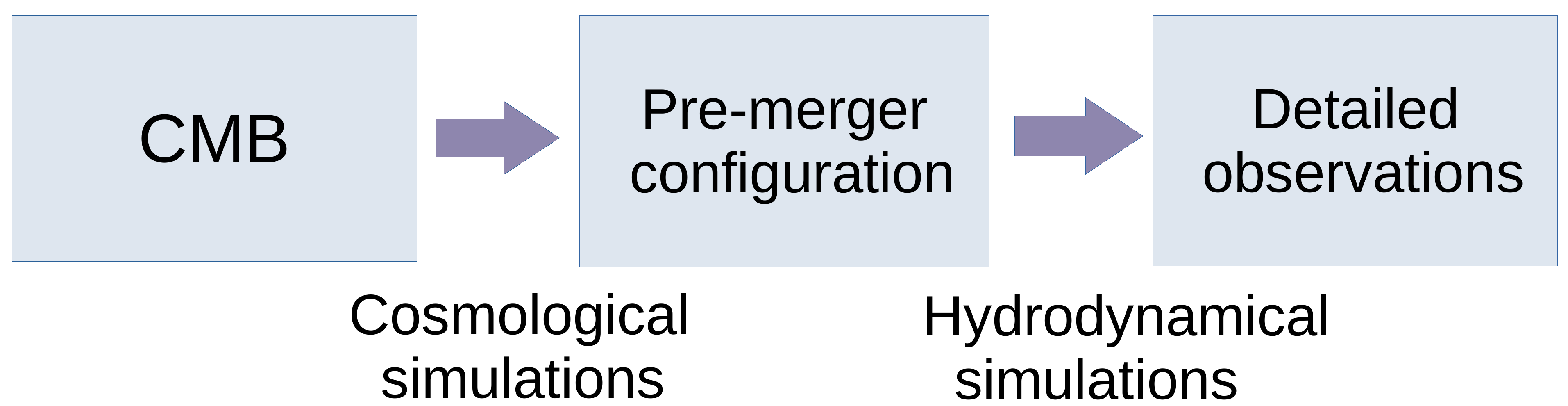}
	\caption{Diagram summarizing the two-step logic of this work: the baryon acoustic oscillations in the cosmic microwave background (CMB) constrain the cosmological parameters in a $\Lambda$CDM universe, fixing the abundance and evolution of its matter content. This information is used in large $N$-body cosmological simulations that allow us to estimate the distribution of structures at any epoch, in particular the cluster pairwise mass function. If $\Lambda$CDM is correct, the distribution at $z = 1$ contains a pair in a pre-merger configuration that, when advanced with a hydrodynamical simulation, reproduces the detailed properties of El Gordo. Such a detailed analysis is not possible without hydrodynamics, while including it precludes the use of a very large simulation volume and the associated statistical power. Previous studies like \citet{Zhang_2015} obtained a good fit to El Gordo using hydrodynamical simulations, but did not quantify whether the required pre-merger configuration is plausible in the surveyed region if the observed anisotropies in the CMB are evolved to $z = 1$ in a $\Lambda$CDM context.}
	\label{Logic_diagram}
\end{figure}

In this paper, we conduct a rigorous analysis to find the probability that a pair of progenitor galaxy clusters at $z \approx 1$ could have given rise to an object like El Gordo within the surveyed region if $\Lambda$CDM is the correct cosmological model. The overall logic of the project is summarized in Figure~\ref{Logic_diagram}. For the second step, we make use of the results obtained by \citet{Zhang_2015} in their hydrodynamical simulations of El Gordo. The main purpose of our work is to check how often the initial conditions of their plausible models arise in a very large $\Lambda$CDM cosmological simulation with side length $6 \, h^{-1}$~cGpc developed by the Juropa Hubble Volume Simulation Project \citep{Watson_2013}.\footnote{\url{https://jubilee.ft.uam.es/}} We refer to this as the Jubilee simulation.

The structure of this paper is as follows: we begin by describing how we find analogues to El Gordo (Section~\ref{Finding_analogues}). The pre-merger configuration obtained from hydrodynamical simulations (Section~\ref{Models}) is compared with the properties of galaxy cluster pairs in the Jubilee simulation (Section~\ref{Jubilee}) using appropriate selection criteria (Section~\ref{Selection_criteria}). We then describe our statistical analysis and its results in Section~\ref{Statistics}. We perform the analysis in two different ways: the power-law method (Section~\ref{Power_law_method}), and lightcone tomography (Section~\ref{Lightcone_tomography}). We combine our results with the \citet{Kraljic_2015} analysis of the Bullet Cluster (Section~\ref{Combined_with_BC}). We then discuss our results in Section~\ref{Discussion}, both in a standard context (Section~\ref{Lower_mass}) and with a non-standard extended gravity law (Section~\ref{MOND}). Finally, we summarize our most relevant conclusions in Section~\ref{Conclusions}.


\section{Finding analogues to El Gordo}
\label{Finding_analogues}

\subsection{Hydrodynamical simulations of the merger}
\label{Models}

\citet{Donnert_2014} conducted the first attempt to reproduce the main characteristics of El Gordo with hydrodynamical simulations in a $\Lambda$CDM cosmology. Their simulation used a Hernquist dark matter density profile \citep{Hernquist_1990}, with the gas following the $\beta$-model of \citet{Cavaliere_1978} with $\beta = 2/3$. They inferred the following model parameters for the El Gordo cluster using the \citet{Menanteau_2012} observations: main cluster mass $M_{200,1} = 1.9 \times 10^{15} M_{\odot}$, secondary cluster mass $M_{200,2} = 8.1 \times 10^{14} M_{\odot}$, main cluster radius $R_{200,1} = 2.55$~Mpc, secondary cluster radius $R_{200,2} = 1.925$~Mpc, relative infall velocity $V_{inf} = 2600$~km/s, gas fraction $f_b \left( <r_{200} \right) = 0.17$ for both clusters, and a small impact parameter of $b = 20$~kpc. Their model was implemented with the magnetohydrodynamics-smoothed particle hydrodynamics (MHD-SPH) code \textsc{gadget-3} \citep{Springel_2005, Dolag_2009}. The model was able to reproduce the total X-ray luminosity, the core distance of the clusters, and the observed offset between the X-ray and SZ centroids. But it was not able to properly reproduce the X-ray morphology: only one tail comes out of the peak emission instead of the observed two. \citet{Donnert_2014} attributed this to the lack of substructure in the simulation, and to the possibility that the parent cluster of El Gordo was a highly disturbed system even before it reached the observed configuration.

The hydrodynamical simulations of \citet{Molnar_2015} used a Navarro-Frenk-White \citep[NFW;][]{Navarro_1997} dark matter density profile and a truncated non-isothermal $\beta$-model with $\beta = 1$ for the gas. They also constrained their model parameters with the \citet{Menanteau_2012} observations, but used slightly different values to \citet{Donnert_2014}: $M_{200,1} = 1.13 \times 10^{15} M_{\odot}$, $M_{200,2} = 1.02 \times 10^{15}$ $M_{\odot}$, $R_{200,1} = 2.304$~Mpc, $R_{200,2} = 1.944$~Mpc, $V_{inf} = 2250$~km/s, $f_b \left( <r_{200} \right) = 0.14$ for both clusters, and $b = 300$~kpc. They used the adaptive mesh refinement (AMR) code \textsc{flash} \citep{Fryxell_2000}. Their use of an AMR code instead of an SPH code \citep[as used in][]{Donnert_2014} was motivated by the fact that an SPH code is prone to suppress turbulent mixing of the gas, which can play an important role in El Gordo \citep{Bauer_2012}. With this model, \citet{Molnar_2015} managed to reproduce its two-tailed X-ray morphology, but they obtained a low X-ray luminosity with respect to the observations. The projected distance between the clusters and the offset between their SZ and X-ray centroids were also larger than observed.

The most recent simulation of El Gordo was conducted by \citet{Zhang_2015}, who explored the parameter space in greater detail. They chose the NFW profile for the dark matter and the \citet{Burkert_1995} profile for the gas. Their simulations used both the SPH code and an AMR code as each offers different advantages. In particular, the \textsc{gadget-2} code (SPH based) was used to perform a large number of merger simulations, exploiting the higher computational speed and effective resolution of SPH codes. Once they identified which simulations could be potential candidates to reproduce the El Gordo merger, they used the \textsc{flash} code (AMR based) to resimulate those mergers with a better handling of the shocks, eddies, and fluid instabilities that arise in the merging process. They ran 123 simulations for different parameters to check which gave a better fit to the El Gordo observations of \citet{Menanteau_2012}. Their simulations can be divided into the following two classes of model:
\begin{enumerate}
	\item Model A: extremely energetic head-on collisions. This category includes the \citet{Donnert_2014} and \citet{Molnar_2015} simulations. The parameters derived by \citet{Zhang_2015} that best match the observed features of El Gordo are: $M_{200,1} = 1.3 \times 10^{15} M_{\odot}$, $M_{200,2} = 6.5 \times 10^{14} M_{\odot}$, $R_{200,1} = 1.66$~Mpc, $R_{200,2} = 1.32$~Mpc, $V_{inf} = 3000$~km/s, $f_b \left( <r_{200} \right) = 0.1$ for both clusters, and $b = 300$~kpc. This set of parameters $-$ referred to as `fiducial model A' by the authors $-$ can generate an X-ray surface brightness distribution similar to the observations. It can also reproduce the observed mass density distribution. But the twin-tailed X-ray morphology is smaller and more asymmetric than observed, and only appears when the clusters are at a projected distance of 600~kpc, smaller than the weak lensing analysis of \citet{Jee_2014} which shows the cores are separated by $\approx 700$~kpc. The offset between the SZ and X-ray centroids is also smaller than observed, while there is a lack of extended X-ray emission in the outer region of the merger. From this and the other Model A simulations of \citet{Zhang_2015}, they reached the following conclusions: (a) collisions with smaller impact parameter are more violent; (b) having a smaller radius for the secondary progenitor cluster may lead to an increase of the X-ray emission in the cluster centre, but only one tail is formed; and (c) having a more unequal mass ratio in this scenario would make the collision less violent, preventing the destruction of the primary cluster's gas core and thus leading to the formation of two peaks in the X-ray emission, contradicting observations. These conclusions are in agreement with the results that \citet{Donnert_2014} and \citet{Molnar_2015} obtained from their simulations.

	\item Model B: these simulations include off-centre collisions of two massive clusters with $b \geq 500$~kpc, making the collisions less violent than case A mergers. The parameters that provide the best fit for the observed characteristics of El Gordo are: $M_{200,1} = 2.5 \times 10^{15} M_{\odot}$, $M_{200,2} = 0.7 \times 10^{15} M_{\odot}$, $R_{200,1} = 2.06$~Mpc, $R_{200,2} = 1.35$~Mpc, $V_{inf} = 2500$~km/s, gas fractions $\left( f_{b1},f_{b2} \right) = (0.05, 0.1)$ for the main and secondary clusters, respectively, and $b = 800$~kpc. This is referred to as `fiducial model B'. The simulation of a merger with these parameters reproduces the temperature and X-ray luminosity of El Gordo, as well as a two-tailed X-ray morphology with a closer resemblance to observations than Model A. The projected distance between the two clusters when this structure is observed is 780~kpc, higher than estimated observationally but still closer to the observational value than the model A result. The offset between the SZ and X-ray centroids is also slightly smaller than observed. An aspect of El Gordo that the model could not reproduce is the X-ray emission in the outer region of the merging cluster. In an attempt to correct for this, \citet{Zhang_2015} changed the gas fraction of the main and secondary cluster to 0.11 and 0.12, respectively, while leaving unchanged the other parameters of fiducial model B. This `Extended Model B' reproduces the X-ray extension while also getting an offset between the SZ and X-ray centroids closer to the observed offset. But this changes the distance between the cluster centres to 890~kpc, significantly larger than observed. After carrying out several Model B simulations, the authors concluded that: (a) the relative velocity needed to reproduce the two-tailed X-ray morphology is lower in Model B ($V_{inf} \approx \left( 1500 - 2500 \right)$~km/s rather than $\approx 3000$~km/s); (b) to have a single X-ray peak in a Model B scenario, the gas fraction of the main cluster must be lower than that of the secondary, but if the gas fraction of the secondary is too large, this would lead to the formation of an unrealistic bright gas core in the centre; (c) mergers with a mass ratio $\la 2$ or $\ga 5$ lead to the formation of a more asymmetric X-ray morphology; (d) a smaller core radius of the secondary cluster leads to brighter X-ray emission in the core and a larger gradient in the X-ray emission; and (e) a relatively large ($\approx 2.5 \times 10^{15} M_{\odot}$) main cluster mass is needed to generate the total X-ray luminosity of El Gordo, which is well constrained observationally.
\end{enumerate}

Out of all these models, it is clear that the observed properties of El Gordo are best reproduced in the \citet{Zhang_2015} fiducial Model B. The values chosen in this model reproduce El Gordo's characteristics fairly well, and are also in agreement with weak lensing observations $-$ the total mass $M_{200,T} = 3.2 \times 10^{15} M_{\odot}$ is very consistent with the weak lensing analysis of \citet{Jee_2014}. The mass ratio between the clusters is slightly higher (more unequal) in fiducial Model B than in the weak lensing observations, though still within uncertainties. We expect that detailed modelling of the collision provides a far better guide to the mass ratio. Therefore, our main analysis uses the parameters of Model B in \citet{Zhang_2015} as our reference for the pre-merger configuration of El Gordo. It is also their recommended model, as evidenced by them conducting an Extended Model B to further optimize the fit to observations.\footnote{Since our analysis considers only the total mass of each cluster, it is not sensitive to their gas fractions. Thus, it is not relevant to discuss whether Extended Model B would be preferred over the fiducial Model B.} In Section~\ref{Lower_mass}, we discuss their Model A and rerun some of our analyses for this case.

\subsection{The Jubilee simulation}
\label{Jubilee}

The Jubilee project consists mainly of two large $N$-body simulations: Small Jubilee, a $3.072 \, h^{-1}$~cGpc cubic box with $3072^3$ particles; and Big Jubilee, a $6 \, h^{-1}$~cGpc cubic box with $6000^3$ particles. Both use the $N$-body code \textsc{cubep\textsuperscript{3}m} \citep{Harnois_2013} and were run on the Juropa Supercomputer at the J{\"u}lich Supercomputer Centre \footnote{\url{https://www.fz-juelich.de/portal/EN/Home/}}. These simulations assume a $\Lambda$CDM cosmology with cosmological parameters $\Omega_{m, 0} = 0.27$, $\Omega_{\Lambda, 0} = 0.73$, $h = 0.7$, $\sigma_8 = 0.8$, $n_s = 0.96$, and $\Omega_{b, 0} = 0.044$. The particle mass is $7.49 \times 10^{10} \, h^{-1} M_{\odot}$ \citep[section~2 of][]{Watson_2014b}. The Jubilee project currently only provides access to results of the Big Jubilee simulation, which have been post-processed in different catalogues with the algorithms known as Amiga Halo Finder (\textsc{ahf}), Friends of Friends (FoF), and Spherical Overdensity (SO), with the latter still under construction (for citations and comparisons between these algorithms, see \citealt{Knebe_2011} and \citealt{Watson_2014}). For this project, we choose the \textsc{ahf} catalogue since it provides a more detailed output with more available parameters.

\textsc{ahf} is an AMR code that identifies and hierarchically classifies structures inside an $N$-particle simulation box \citep{Gill_2004, Knollmann_2009}. Its volume is covered with a regular grid of user-defined size. In each cell, the particle densities are calculated using a triangular shaped cloud weighting scheme \citep{Hockney_1988}. If the particle density exceeds a given threshold, the cell will be refined and covered by a finer grid with half the linear cell size. This process is repeated until a grid is reached which needs no further cell refinement. Following this procedure yields a grid hierarchy starting from the finest grid to the coarsest in such a way that isolated regions from the finer grids identified as possible haloes are linked to their corresponding volumes in the coarser grid.\footnote{The volume covered by a fine grid is a subset of the volume covered by the coarser grids.} From this, a tree of nested grids is constructed. In case two patches which are isolated on one level link into the same patch on the next coarser grid, the two branches of the grid tree join. Once this is done for each nest of grids, the grid forest is constructed and the classification of substructure can be made $-$ starting from the coarsest level downward to the finer levels, once the finer level splits up into two or more isolated patches, the patch containing the most particles is chosen as the main branch, while the others are classified as substructures.

The main properties of each halo are its catalogue radius $r_{Jub}$ and catalogue mass $M_{Jub}$, with $_{Jub}$ subscripts denoting values derived from the Jubilee catalogue. This determines the average density $\overline{\rho} \left( <r_{Jub} \right)$ within the radius $r_{Jub}$. We use this to obtain the overdensity
\begin{eqnarray}
	\label{Delta_Jub}
	\Delta_{Jub} ~&\equiv&~ \frac{\overline{\rho} \left( <r_{Jub} \right)}{\rho_c} \, , \text{ where} \\
	\rho_c ~&\equiv&~ \frac{3H^2}{8\mathrm{\pi}G}
\end{eqnarray}
is the critical density of the universe at the snapshot redshift, when the Hubble constant is $H$. In the Jubilee catalogues, $\Delta_{Jub} = 178$. For a few cases, we checked that the values of $\Delta_{Jub}$ listed in the catalogue correspond to our calculations using Equation \ref{Delta_Jub}. In the catalogue files, distances are in cMpc/$h$, masses in $M_{\odot}/h$, and peculiar velocities in km/s, where $h$ is the present Hubble constant $H_0$ in units of 100~km/s/Mpc. We assume $h = 0.7$ throughout this article for consistency with the Jubilee simulations. We convert co-moving quantities to physical for our analysis, requiring us to add the Hubble flow velocity to the peculiar velocities listed in the catalogue.

The overdensity $\Delta_{Jub}$ used in the Jubilee catalogues is not the standard value of 200. For comparison with constraints derived from hydrodynamical simulations (Section~\ref{Models}), we convert the Jubilee quantities using the procedure described below. We define the scaled density within the radius $r_{Jub}$ as $\widetilde{\rho} \equiv \Delta_{Jub}/200$, where we use $\widetilde{q}$ to denote $q/q_{200}$ for any quantity $q$ with value $q_{200}$ when defined within the radius $r$ for which $\overline{\rho} \left( <r \right) = 200 \, \rho_c$. We guess the scaled catalogue mass $\widetilde{m}$ and thereby determine
\begin{eqnarray}
	\widetilde{r} ~=~ \sqrt[3]{\frac{\widetilde{m}}{\widetilde{\rho}}} \, .
\end{eqnarray}
To make further progress, we assume the cluster follows the NFW profile \citep{Navarro_1997} with concentration parameter $c$.\footnote{$r_{200} \equiv c r_s$, where $r_s$ is the radial scale of the profile.} We use equation 4 of \citet{Duffy_2008} to get that:
\begin{eqnarray}
	c ~=~ 6.71 a^{0.44} \left( \frac{M_{200}}{2 \times 10^{12} h^{-1} M_\odot} \right)^{-0.091} \, ,
\end{eqnarray}
where $a \equiv 1/\left(1 + z\right)$ is the cosmic scale factor. For a NFW profile, we must have that:
\begin{eqnarray}
	\widetilde{m} \left[ \ln \left( 1 + c \right) - \frac{c}{1 + c} \right] ~=~ \ln \left( 1 + c\widetilde{r} \right) - \frac{c\widetilde{r}}{1 + c\widetilde{r}} \, .
\end{eqnarray}
To ensure that this equation is satisfied, we set the difference between its left and right sides to 0 by varying our guess for $\widetilde{m}$ using the Newton-Raphson algorithm. In this way, we obtain $m_{200}$ and $r_{200}$ for use in subsequent analyses. We call these quantities the virial mass and radius, respectively.

\subsection{Selection criteria}
\label{Selection_criteria}


Our main goal is to find galaxy cluster pairs on an orbit similar to that of the El Gordo progenitors, whose properties we obtain using the hydrodynamical simulations discussed in Section~\ref{Models}. To avoid contamination from galaxies, we restrict our sample to only those haloes with $M_{200} \geq 3.5 \times 10^{13}$ $M_{\odot}$. This leaves us with 4129462 haloes, which we assume are all in the galaxy cluster regime.

Comparing all of these haloes with each other would be extremely computationally expensive. To improve the efficiency of our pair-finding algorithm, we only consider the region within some distance $r_{max}$ of a halo with virial mass $M$, where
\begin{eqnarray}
	r_{max} ~=~ r_{_0} \sqrt[3]{\frac{M}{M_0}} \, ,
\end{eqnarray}
with $r_{_0} = 40$~Mpc and $M_0 = 10^{13}$ $M_{\odot}$. We can estimate the orbital period of any pair by applying Kepler's laws. Since $r_{max} \propto M^{1/3}$, the period is independent of $M$, allowing us to calculate it for a cluster with mass $M_0$. In this case, the minimum possible period $P_0$ arises if the total mass takes the highest possible value of $4.6 \, M_0$ (Equation \ref{Mass_ratio_limit}), the apocentre is $r_{_0}$, and the pericentre is 0, making the semi-major axis $r_{_0}/2$. For this configuration,
\begin{eqnarray}
	P_0 ~=~ 2\mathrm{\pi} \sqrt{\frac{\left(\frac{r_{_0}}{2}\right)^3}{4.6 \, G M_0}} ~=~ 1220 \, \text{Gyr.}
\end{eqnarray} 
Thus, it is clear that we do not miss any pairs which could have turned around from their initial expansion and subsequently reached pericentre within the lifetime of the Universe. Nonetheless, our restriction on the maximum separation of each pair greatly reduces the computational cost of our algorithm as the simulation box size is $\left( 6 \, h^{-1} \, \text{cGpc} \right)^3$.

For a pair to be analogous to El Gordo, it should also have turned around from the cosmic expansion (Section~\ref{Turnaround_condition}), have an appropriate total mass and mass ratio (Section~\ref{Mass_condition}), infall velocity (Section~\ref{Infall_velocity_condition}), redshift (Section~\ref{Redshift_condition}), and impact parameter (Section~\ref{Impact_parameter_condition}). We discuss these criteria next, but usually omit the last condition to be conservative.

\subsubsection{Requiring turnaround}
\label{Turnaround_condition}

Amongst the pairs with separation $< r_{max}$, we consider only those that have turned around from the cosmic expansion to avoid e.g. two clusters separated by 30~Mpc with velocities close to the Hubble flow. In other words, the pairwise relative separation $\bm{r}_{rel}$ and relative velocity $\bm{v}_{rel}$ must satisfy
\begin{eqnarray}
	\bm{r}_{rel} \cdot \bm{v}_{rel} ~<~ 0 \, .
	\label{Turnaround_equation}
\end{eqnarray}
This restricts our sample to cluster pairs which may subsequently undergo a close interaction reminiscent of El Gordo. Since most of the time in an orbit is spent close to apocentre, we should easily be able to find El Gordo analogues in this manner if they are common in the Jubilee simulation. A very small number of analogues could be missed because e.g. at the time of the snapshot, an interacting pair is caught shortly after pericentre. However, given the significant amount of dynamical friction expected in an interaction like El Gordo, we expect this to be a fairly short-lived phase compared to the pre-merger phase. Even if a cluster pair is after pericentre at the snapshot time, it could still satisfy Equation \ref{Turnaround_equation} provided the pair has turned around for a second time. This may be appropriate to the case of El Gordo \citep{Karen_2015}. Our analysis also misses El Gordo progenitors that will turn around after the snapshot time. However, the time required to turn around and reach pericentre means we would need to find such pairs at a much higher redshift than the $z = 0.87$ of El Gordo. It would also be very difficult to know for sure whether a pair will indeed turn around subsequent to the snapshot. In what follows, we will assume that only a small fraction of possible El Gordo progenitors are missed by imposing the turnaround condition.

\subsubsection{Mass}
\label{Mass_condition}

The fact that El Gordo is an interaction between similar mass clusters is expected to significantly lower the predicted number of analogous pairs in the survey region. This is because minor mergers are much more common than major mergers. According to Model B of \citet{Zhang_2015}, the mass ratio of the El Gordo progenitors was 3.6. Therefore, we require each galaxy cluster pair to satisfy
\begin{eqnarray}
	\frac{m_{max}}{m_{min}} ~\leq~ 3.6 \, ,
	\label{Mass_ratio_limit}
\end{eqnarray}
where $m_{min}$ and $m_{max}$ are the virial masses of the member with the lower and higher mass, respectively.

The total mass $M$ of El Gordo is one of the most important parameters. Model B of \citet{Zhang_2015} implies that $M = 3.2 \times 10^{15} M_\odot$, so we require
\begin{eqnarray}
	\widetilde{M} ~\equiv~ \log_{10} \left( \frac{m_{min} + m_{max}}{M_\odot} \right) ~>~ \widetilde{M}_{EG} = 15.50 \, .
	\label{M_tilde}
\end{eqnarray}
This condition leaves us with no analogous systems in the entire Jubilee volume, so our statistical analysis (Section~\ref{Statistics}) is based on how the number of analogues changes as the limit on $\widetilde{M}$ is increased from a much lower value. The results are extrapolated up to the El Gordo value of 15.50, allowing us to obtain reliable statistics even though $\ll 1$ analogous pair exists in the Jubilee volume. The accuracy of this approach is discussed further in Section~\ref{Jubilee_reliability}.

\subsubsection{Infall velocity}
\label{Infall_velocity_condition}

To enable a comparison between galaxy cluster pairs caught at different phases of their orbit in a Jubilee snapshot, we use energy conservation under a point mass Newtonian potential to determine the relative velocity $v \left( 2 R_T \right)$ when the clusters are separated by $2 R_T$, where $R_T$ is the sum of their virial radii. We then define a scaled velocity $\widetilde{v}$ to facilitate a fair comparison between cluster pairs with a very different total mass.
\begin{eqnarray}
	\widetilde{v} ~\equiv~ v \left( 2 R_T \right) \div \overbrace{\sqrt{\frac{GM}{R_T}}}^{v_{esc}} \, ,
	\label{v_tilde}
\end{eqnarray}
where $M$ is the sum of the cluster virial masses. The idea is to scale the relative velocity to the escape velocity $v_{esc}$. We set $\widetilde{v} = 0$ for a pair with insufficient energy to reach a separation of $2 R_T$. To match the observed properties of El Gordo, we require that $\widetilde{v} > \widetilde{v}_{EG} = 1.24$ for our nominal Model B. If instead we use Model A, $\widetilde{v}_{EG}$ becomes 1.77. Although a detailed analysis of the Bullet Cluster is beyond the scope of this work, for comparison we use table~2 of \citet{Lage_2014} to estimate that its $\widetilde{v}_{BC}$ is only 0.80.


\subsubsection{Redshift}
\label{Redshift_condition}

The observed merging configuration of El Gordo is viewed at $z = 0.870$. However, we expect our procedure to find possible progenitor pairs before they interact. This is because:
\begin{enumerate}
	\item it is difficult for the position-based \textsc{ahf} \citep{Knollmann_2009} to distinguish closely interacting pairs, and
	\item most of the time in an orbit is spent near apocentre.
\end{enumerate}
Thus, we can only explain the observed configuration of El Gordo if we find progenitors at a slightly higher redshift. For this, the temporally closest snapshot available in the Jubilee simulation is at $z = 1$. The time difference between $z = 1$ and $z = 0.870$ is $\Delta t = 559$~Myr.

To check if this time lag suits the aforementioned purpose, we compare it with the pre-merger timeframes obtained with hydrodynamical simulations of El Gordo. \citet{Zhang_2015} estimate that its observed state corresponds to 140~Myr after pericentre. The total time lag should be much larger as clusters spend most of their time going from apocentre to pericentre.

\begin{figure}
	\centering
	\includegraphics[width = 8.5cm] {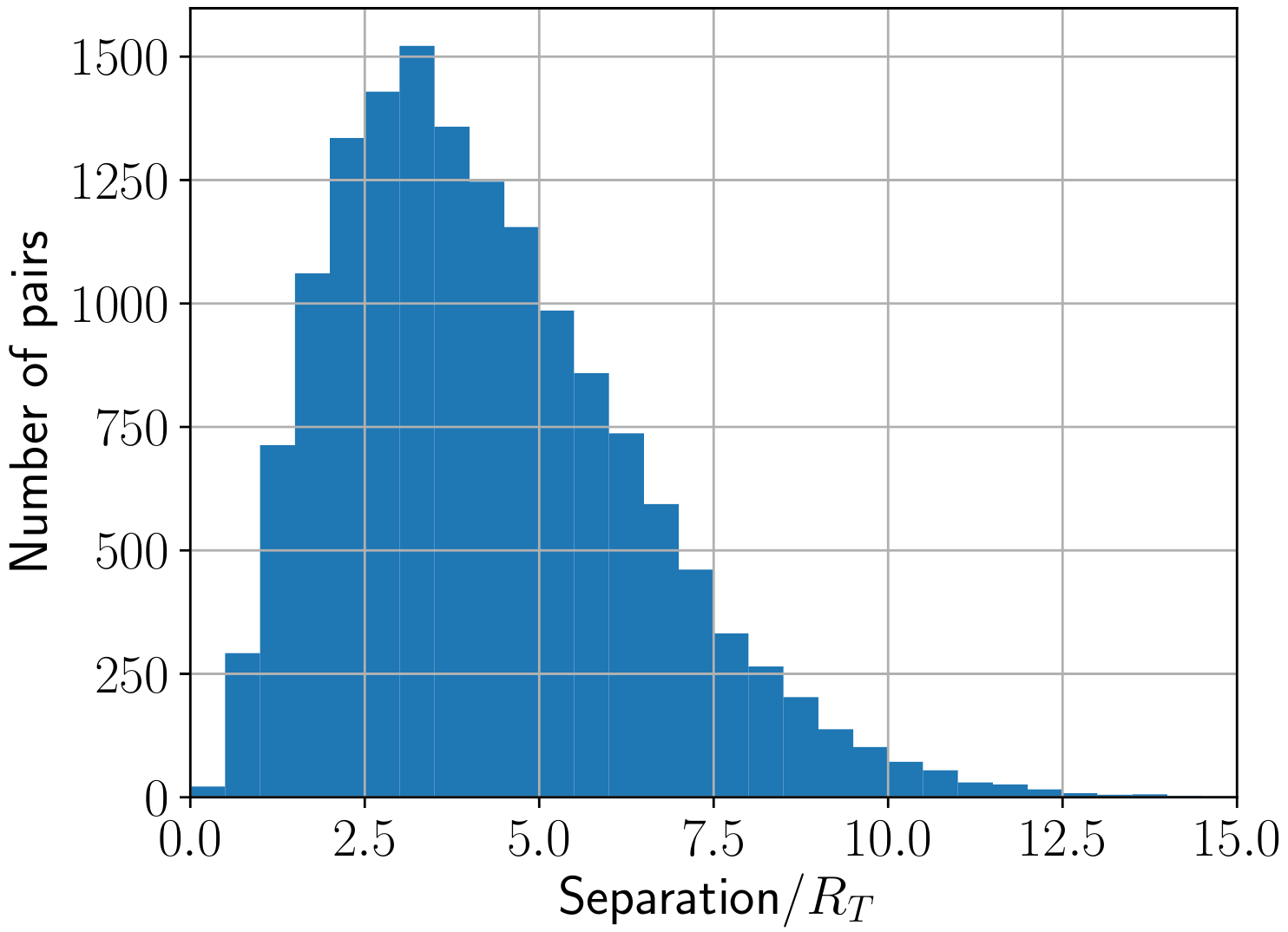}
	\caption{Histogram of the ratio $\widetilde{d}$ between the separation of candidate El Gordo analogue pairs and the sum of their virial radii. The pairs shown here are identified at $z = 1$ and satisfy the turnaround, mass ratio, and $\widetilde{v}$ conditions, but have a smaller total mass than El Gordo (no pairs would be left otherwise). Most pairs have $\widetilde{d} > 4$. There are almost no pairs with $\tilde{d} < 0.5$, and only 314/15035 have $\tilde{d} < 1$.}
	\label{hist_dist}
\end{figure}

To get an idea of how long the clusters spend before pericentre, we assume that they are in free-fall at their escape velocity ($\widetilde{v} = 1$) and that their initial separation $r = R_T$, since it would be difficult for \textsc{ahf} to distinguish clusters with overlapping virial volumes. Assuming the clusters are point masses of $M_{200}$, the free-fall time $t_{ff}$ is
\begin{eqnarray}
	t_{ff} ~=~ \frac{r^{3/2}\sqrt{2} }{3\sqrt{GM_{200,T}}} ~=~ 748 \, \text{Myr} \, .
	\label{ff_eq}
\end{eqnarray}
Since $\widetilde{v}$ is actually 1.24, a better estimate would be if we reduce $t_{ff}$ by this factor, yielding 603~Myr. Adding this to the post pericentre time lag of 140~Myr given in table~2 of \citet{Zhang_2015}, we obtain a total time lag of $\Delta t = 743$~Myr. Thus, our estimate of 559~Myr is conservative even if all our candidate pairs in the Jubilee simulation have a separation of $R_T$. However, the actual separations are typically much larger (Figure~\ref{hist_dist}). This is partly because clusters spend relatively little time near pericentre, but also due to the difficulties of \textsc{ahf} in resolving clusters which are merging. Other halo finders such as \textsc{rockstar} \citep {Behroozi_2013} use velocity data to alleviate this problem. This is not the case with \textsc{ahf}, so the pairs we identify must be more widely separated. According to Equation \ref{ff_eq}, this implies a much longer $\Delta t$.

Other studies also obtain a significantly larger time lag than estimated above. \citet{Donnert_2014} used an initial time ($t = 0$) when the cluster separation is slightly above the sum of their virial radii (see their figure~4). They found that the observed configuration of El Gordo takes place at $t = 1750$~Myr, shortly after a first passage through pericentre. \citet{Karen_2015} propose two different scenarios that could describe today's observations of El Gordo: the clusters could either be right after their first passage through pericentre \citep[as suggested by][]{Donnert_2014}, or they could be returning to pericentre for a second encounter. The first case corresponds to a time since pericentre of $t_{out} = 460$~Myr, while the second scenario $-$ favoured in the author's conclusions $-$ corresponds to a time since pericentre of $t_{ret} = 910$~Myr. This exceeds our estimated $\Delta t$ even without any time lag before pericentre.

Leaving a longer time lag between the Jubilee and observed configurations would mean choosing a higher redshift for our analysis. This lowers the chance of finding an El Gordo analogue as the number of fast collisions of massive objects decreases with increasing redshift. Therefore, we conservatively adopt $z = 1$ as the redshift at which we search for El Gordo analogues. In reality, the time lag could be significantly larger, so a higher $z$ may be more appropriate.

\subsubsection{Impact parameter}
\label{Impact_parameter_condition}

To get a sufficiently strong interaction between the clusters, their impact parameter should be rather small. To calculate this for cluster pairs in the Jubilee simulation, we begin by using energy conservation to estimate their relative velocity $v_\infty$ at large separation. For pairs without sufficient energy to escape, we record $\widetilde{b} = 0$. Otherwise, we calculate
\begin{eqnarray}
	\widetilde{b} ~\equiv~ \frac{\left| \bm{r}_{rel} \times \bm{v}_{rel} \right|}{v_\infty R_T} \, .
	\label{b_tilde}
\end{eqnarray}
Model B of \citet{Zhang_2015} implies that $\widetilde{b}_{EG} = 3.42$, which is appropriate for an off-centre collision. We do not directly include the requirement of a lower $\widetilde{b}$ to be more conservative. This condition should exclude only a small proportion of pairs (Section~\ref{Power_law_method}).

\section{Statistical analysis and results}
\label{Statistics}

We fit the mass distribution of pairs analogous to El Gordo according to some subset of the criteria discussed in Section~\ref{Selection_criteria}. The fitting procedure is described in Section~\ref{Mass_function_fitting}. These fits are used in two methods to quantify the likelihood of observing El Gordo within the surveyed region: the power-law method (Section~\ref{Power_law_method}) and lightcone tomography (Section~\ref{Lightcone_tomography}). In both cases, the use of fitting functions is necessary because the mass of El Gordo is larger than the most massive pair to satisfy the other conditions. However, we will see that it is only necessary to extrapolate $\widetilde{M}$ by $\approx 0.6$ dex. Section~\ref{Combined_with_BC} discusses the implications of our results in combination with the Bullet Cluster.

\subsection{Fitting the mass function of El Gordo-like pairs}
\label{Mass_function_fitting}

\begin{figure}
	\centering
	\includegraphics[width = 8.5cm] {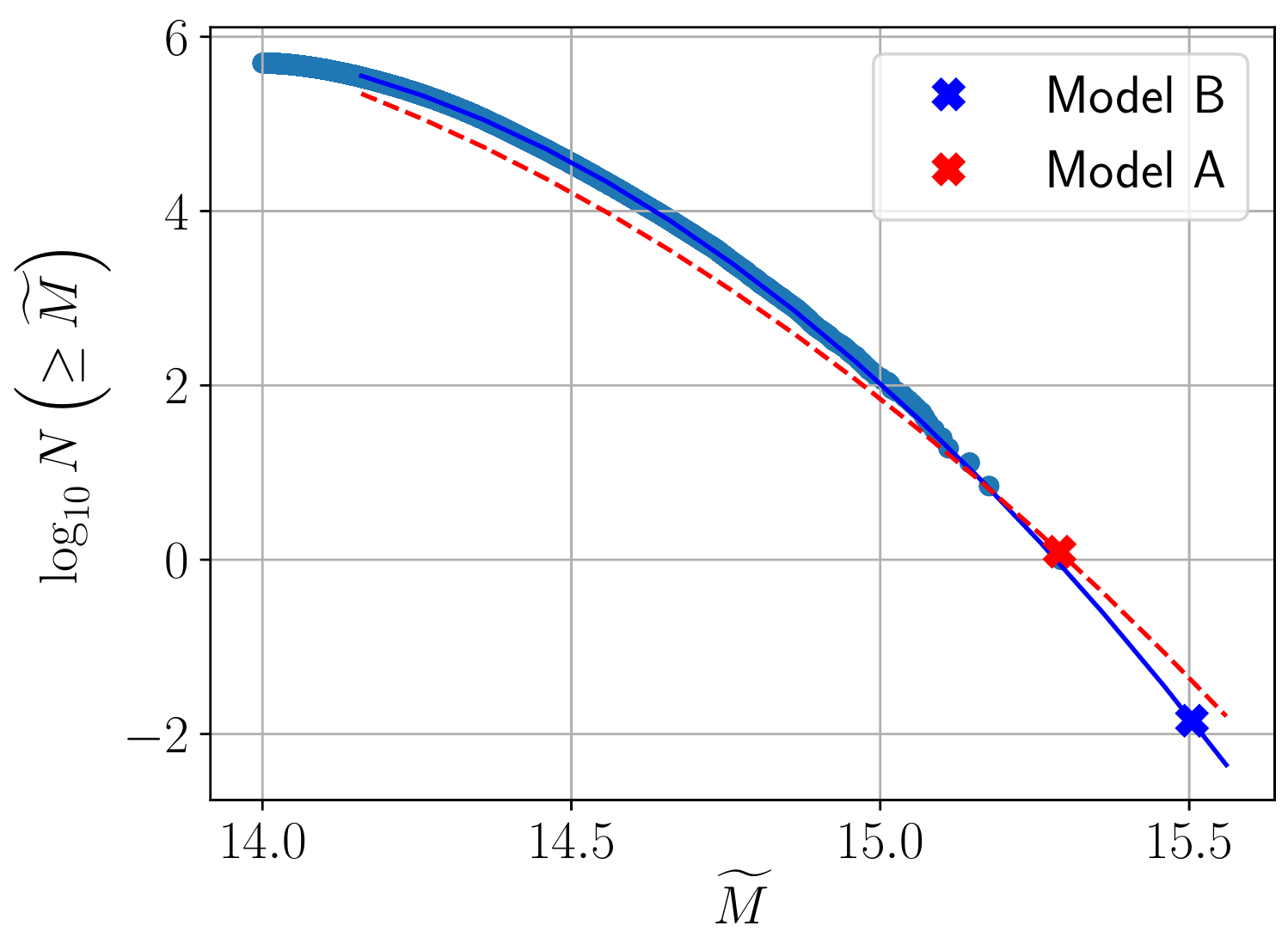}
	\caption{The blue data points show the cumulative $\widetilde{M}$ distribution for haloes in the Jubilee simulation at $z = 1$ that have turned around from the cosmic expansion and have mass ratio $\leq 3.6$ (Model B). The blue solid line is the quadratic fit to this distribution. The red dashed line is the analogous fit when we instead require a mass ratio $\leq 2$ (Model A). For clarity, individual points are not shown in this case. The crosses indicate the $\widetilde{M}$ of each model.}
	\label{cumul_mass}
\end{figure}

With the 499285 pairs that pass the mass ratio and turnaround conditions, we sort the pairs into ascending order of total mass as quantified by $\widetilde{M}$ (Equation \ref{M_tilde}). This lets us obtain the cumulative distribution of $\widetilde{M}$, i.e. the number of pairs $N \left( \geq \widetilde{M} \right)$ whose $\widetilde{M}$ equals or exceeds some value. We show this distribution in Figure~\ref{cumul_mass} for both Model A and Model B parameters. There is a slight difference caused by the smaller mass ratio in Model A, which in the previous step reduces the number of candidates to 379612. We adopt the Model B values as nominal, but it is also helpful to consider Model A because its lower mass for El Gordo is less problematic. We discuss Model A in more detail in Section~\ref{Lower_mass}.

\begin{figure}
	\centering
	\includegraphics[width = 8.5cm] {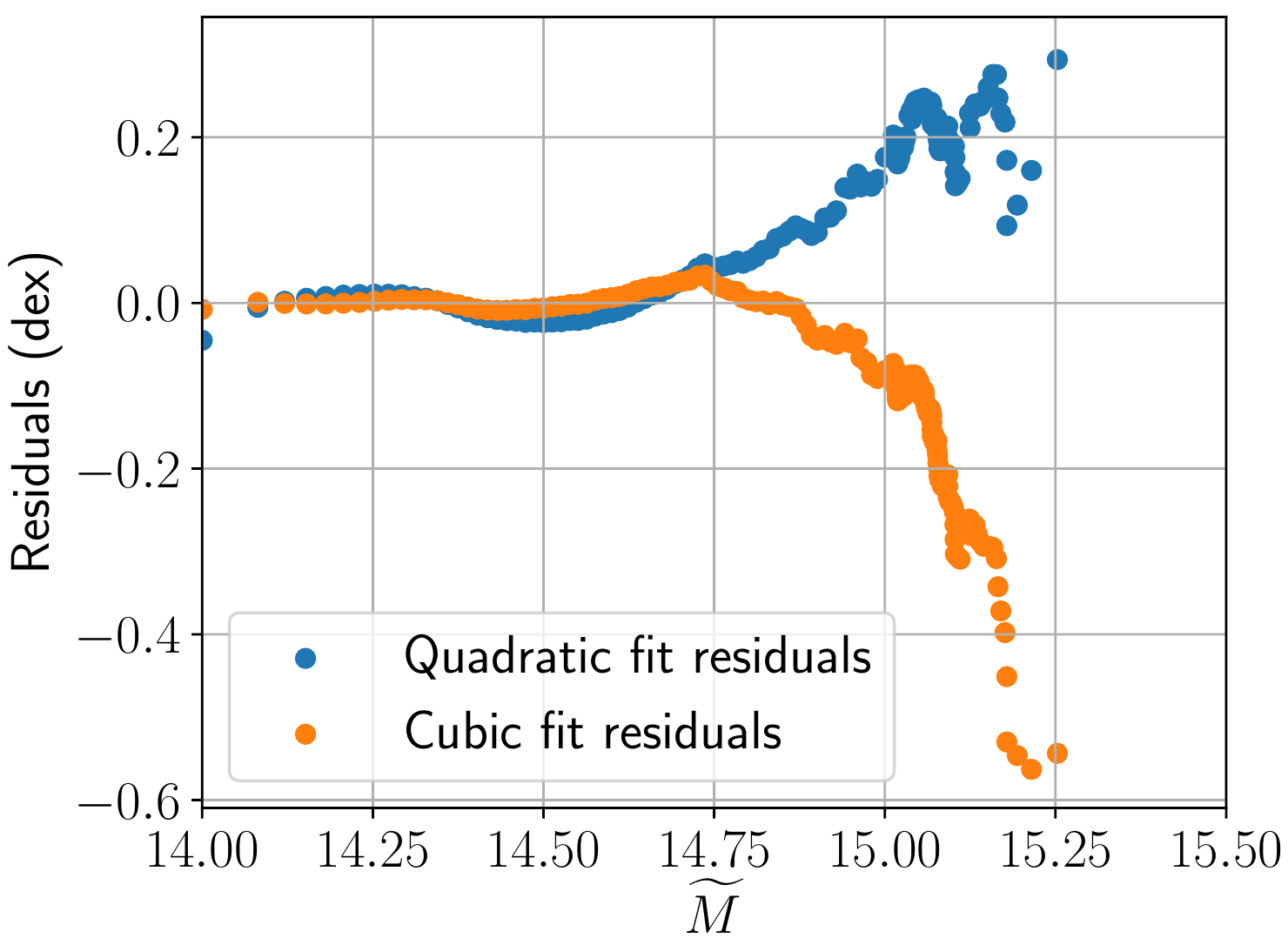}
	\caption{Residuals of the quadratic and cubic fits to the cumulative mass distribution for turned around halo pairs in the Jubilee simulation at $z = 1$ whose mass ratio $\leq 3.6$.}
	\label{residuals}
\end{figure}

By fitting the mass distribution analytically, we can extrapolate the number of pairs analogous to El Gordo which have $\widetilde{M} > \widetilde{M}_{EG}$, which is the more problematic side for $\Lambda$CDM. Therefore, choosing the correct fitting function for the mass distribution is critical to achieving an accurate estimate of how unlikely the observation of El Gordo is in a $\Lambda$CDM context. We initially choose a quadratic fit of $\log_{10} N \left( \geq \widetilde{M} \right)$ against $\widetilde{M}$ since this seems better suited to the distribution. This provides a very good fit (Figure~\ref{cumul_mass}). To check if it can be improved with a higher order fit, we compare the residuals to those of a cubic fit (Figure~\ref{residuals}). It is clear that a cubic would overfit the distribution and provide a poorer match to Jubilee data at the critically important high mass end. Therefore, we use a quadratic fitting function in the remainder of this article. The cumulative $\widetilde{M}$ distribution of pairs that pass the mass ratio, turnaround, and $\widetilde{v}$ conditions is
\begin{eqnarray}
	\log_{10} N \left( \geq \widetilde{M} \right) ~=~ c_0 + c_1 \widetilde{M} + c_2 \widetilde{M}^2 \, ,
	\label{Mass_function_fit}
\end{eqnarray}
with the fitting coefficients listed in Appendix \ref{Appendix} for the Jubilee snapshots at $z = 0$, 0.509, and 1. We use these fits to quantify the probability of finding analogues to the presumed El Gordo progenitors at $z = 1$. This is done using two analysis techniques, which we describe next.

\subsection{Power-law method}
\label{Power_law_method}

Once we know the effective volume $V_{eff}$ of the survey which discovered El Gordo, we can compare this to the Jubilee simulation volume to determine the expected number of El Gordo analogues in the survey volume. From this, we can quantify the $P$-value for a $\Lambda$CDM cosmology of the fact that El Gordo was discovered with its observed properties in the surveyed region. The basic principle is to find the chance that this contains a cluster collision with equally or even more extreme properties.

We extrapolate our fit to the cumulative pairwise mass distribution to get the number of El Gordo analogues in the $ \left( 6 \, h^{-1} \, \text{cGpc} \right)^3$ simulation volume at $z = 1$. By repeating this at $z = 0.509$, we obtain the growth index
\begin{eqnarray}
	k ~\equiv ~ \frac{\Delta \ln n}{\Delta \ln a} \, ,
	\label{k}
\end{eqnarray}
where $n$ is the number of El Gordo analogues in the simulation volume, and the finite difference is taken between the Jubilee snapshots at $z = 1$ and $z = 0.509$. We avoid using the $z = 0$ data because structure formation is slowed down at low $z$ by the effect of dark energy, but its impact should not be too significant at $z \ga 0.5$. When fitting the mass function using the widest available range ($\widetilde{M} \geq 14.0$), we obtain that $k = 24.81$ for the $\widetilde{v} > 0$ condition and $k = 35.55$ for the more realistic $\widetilde{v} > \widetilde{v}_{EG}$ condition.

We then use $k$ to determine $V_{eff}$. The main idea is that since $k \gg 1$, the effective survey depth is limited by the fact that analogues to El Gordo rapidly become very rare with increasing $z$. To find the number of El Gordo analogues at $a < 0.5$ along our past lightcone, we should ideally integrate $n$ over $a \leq 0.5$ while accounting for the variation of $n$ with $a$. For $k \gg 1$, this is approximately equivalent to neglecting the redshift dependence of $n$ and only integrating out to e.g. $a = 0.45$. Since the co-moving number density of El Gordo analogues increases with $a$ as $n \appropto a^k$ with $k$ given by Equation \ref{k}, the effective `depth' of the survey is
\begin{eqnarray}
	\Delta a ~\approx~ \frac{a}{k + 1} \, ,
\end{eqnarray}
where $a = 0.5$. In practice, this is not exactly correct since the co-moving volume $V_c$ per unit $a$ along our past lightcone satisfies
\begin{eqnarray}
	\frac{dV_c}{da} ~=~ \frac{c A {d_c}^2}{a H} \, ,
	\label{dVc_da}
\end{eqnarray}
where $A$ is the sky area of the survey in natural units (steradians), $H$ is the Hubble constant at epoch $a$, and $d_c$ is the co-moving radial distance to an object at that epoch.
\begin{eqnarray}
	d_c ~\equiv~ \int_{t_i}^{t_0} \frac{c \, dt}{a} \, .
\end{eqnarray}
The integral must be taken up to the present time $t_0$ from the time $t_i$ corresponding to the epoch at which we wish to know $d_c$. If this is the epoch when $a = 0.5$, we have that $H \appropto a^{-1.16}$ while $d_c \appropto a^{-1.50}$, so $dV_c/da \appropto a^{-2.83}$. This means that the number of analogues per unit $a$ along the past lightcone is $\appropto a^{k - 2.83}$, so the effective survey depth is
\begin{eqnarray}
	\Delta a ~=~ \frac{a}{k - 1.83} \, .
	\label{Delta_a}
\end{eqnarray}
Since we know the sky coverage of the survey, its effective co-moving volume is then
\begin{eqnarray}
	V_{eff} ~=~ A {d_c}^2 \Delta d_c \, ,
	\label{V_eff}
\end{eqnarray}
which gives $V_{eff} =  6.33 \times 10^8$~cMpc\textsuperscript{3} ($4.17 \times 10^8$~cMpc\textsuperscript{3}) when we require $\widetilde{v} > 0$ ($\widetilde{v} \geq \widetilde{v}_{EG}$). To get the co-moving volume in the past lightcone between $a$ and $a - \Delta a$, we need the difference in $d_c$ between these epochs, which we call $\Delta d_c$. We have simplified our equations by exploiting the fact that the angular diameter distance is $d_c/a$.

\begin{figure}
	\centering
	\includegraphics[width = 8.5cm] {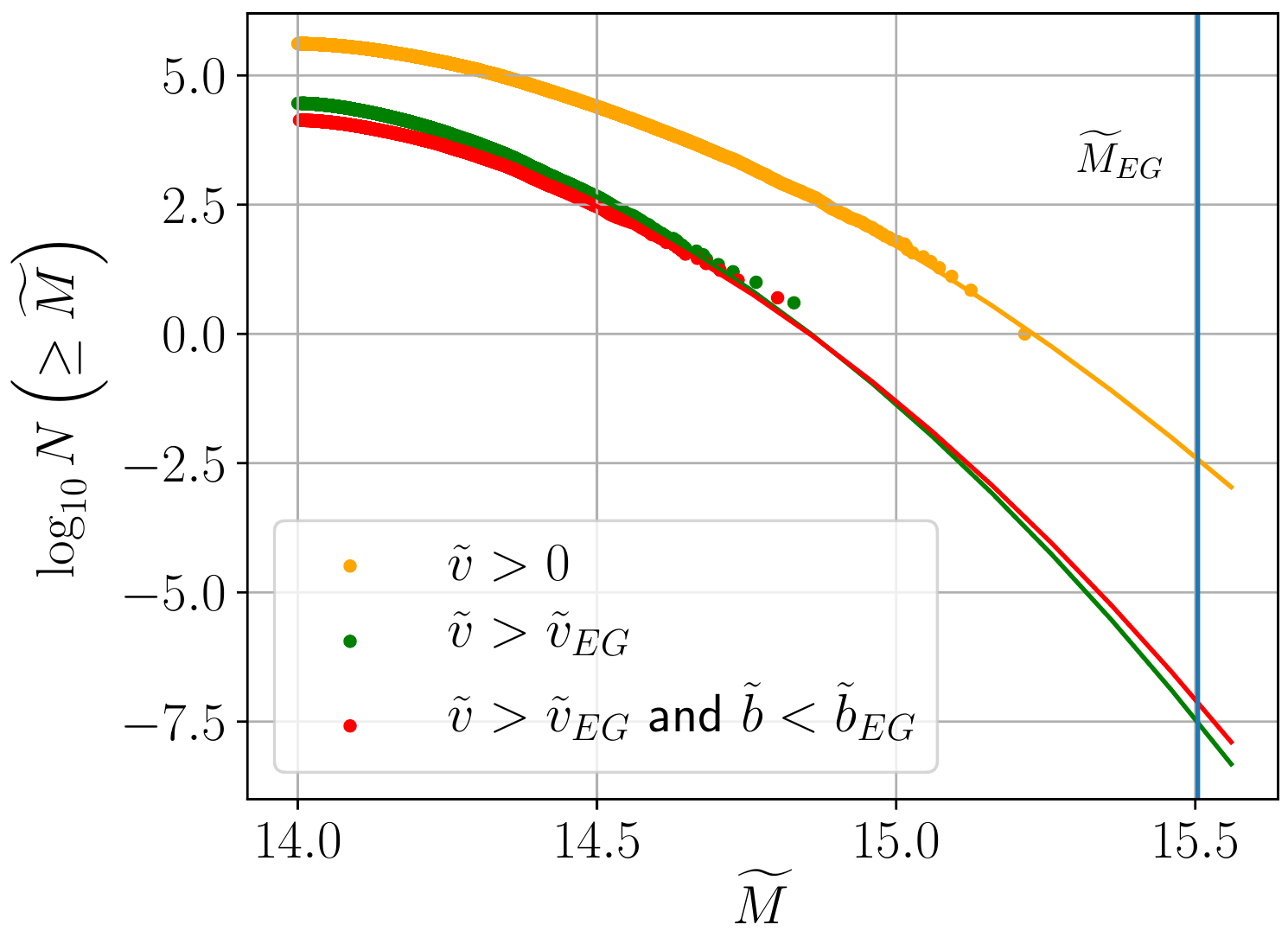}
	\caption{The cumulative $\widetilde{M}$ distribution with the conditions $\widetilde{v} > 0$ (orange), $\widetilde{v} > \widetilde{v}_{EG}$ (green), and $\widetilde{v} > \widetilde{v}_{EG}$ plus $\widetilde{b} < \widetilde{b}_{EG}$ (red). The blue vertical line marks the El Gordo $\widetilde{M}$.}
	\label{allcond}
\end{figure}

The basic characteristics of El Gordo were constrained for the first time in a survey with area $A = 755 \, deg^2$ \citep{Menanteau_2012}. As discussed in Section~\ref{Selection_criteria}, we also add several conditions to obtain a cumulative mass distribution for El Gordo-like pairs. We add these in stages to see how the number of analogues decreases as each condition is imposed. The conditions are: $\widetilde{v} > 0$, $\widetilde{v} \geq \widetilde{v}_{EG}$, and $\widetilde{v} \geq \widetilde{v}_{EG}$ plus $\widetilde{b} \leq \widetilde{b}_{EG}$. Our results are shown in Figure~\ref{allcond}.

Since we only consider clusters with $m/M_{\odot} \geq 3.5 \times 10^{13} \approx 10^{13.5}$, there are no pairs less massive than $\widetilde{M} = 13.5 + \log_{10} \left( 2 \right) = 13.80$. Edge effects are thus very significant at this mass. To avoid edge effects at the low mass end, from now on we only consider pairs with $\widetilde{M} > 13.5 + \log_{10}(4.6) = 14.16$ unless explicitly stated otherwise, allowing for the maximum allowed mass ratio of 3.6 (Equation \ref{Mass_ratio_limit}). Ideally, we would impose a much higher floor and consider only those cluster pairs with similar mass to El Gordo. However, such objects are very rare, so fitting only the very high mass end can lead to an inaccurate estimation of $n$ (see Figure~\ref{allcond}). This effect is more pronounced when we add the $\widetilde{v} \geq \widetilde{v}_{EG}$ condition because in the high mass bin most relevant to El Gordo, the number of pairs starts decreasing very rapidly with $\widetilde{v}$ once $\widetilde{v} \ga 0.6$ (Section~\ref{Lower_mass}). Therefore, our main results rely on all pairs with $\widetilde{M} > 14.16$, maximising the amount of Jubilee data used. This means our results rely on the top $40-50\%$ of our pairwise mass distribution.

We use the survey volume $V$ to find the expected number of analogues $N_{EG}$ within it. Since $N_{EG} < 1$, a situation `as bad as or worse than' reality for $\Lambda$CDM implies in this case that the survey region contains $\geq 1$ El Gordo-like system. In a $\Lambda$CDM context, the probability of this occurring is given by standard Poisson statistics \citep{Watson_2014}.
\begin{eqnarray}
	P ~=~ 1 - \exp \left( -N_{EG} \right) \, .
\end{eqnarray}
Once we have the $P$-value, we can express it as an equivalent number of standard deviations for a Gaussian distribution, which we call the $\chi$ value. By definition, $P$ and $\chi$ are related to each other by the Gaussian distribution.
\begin{eqnarray}
	1 - \frac{1}{\sqrt{2 \mathrm{\pi}}} \int_{-\chi}^\chi \exp \left( -\frac{x^2}{2} \right) \, dx ~\equiv~ P \, .
\label{P-chi}
\end{eqnarray}
In this contribution, we will usually convert a $P$-value into the statistical significance $\chi$ by solving this equation using the Newton-Raphson algorithm. The nominal $\chi$ values that we obtain from this analysis are $\chi = 4.61$ for the $\widetilde{v} > 0$ condition and $\chi = 6.69$ for the more realistic $\widetilde{v} \geq \widetilde{v}_{EG}$ condition. Since this already exceeds the $5\sigma$ threshold, we do not show results for imposing both $\widetilde{v} \geq \widetilde{v}_{EG}$ and $\widetilde{b} \leq \widetilde{b}_{EG}$. To illustrate the effect of the $\widetilde{b}$ condition, we show how the $z = 1$ mass function changes if we additionally impose this condition (Figure~\ref{allcond}). Requiring that $\widetilde{b} < \widetilde{b}_{EG}$ only slightly reduces the number of analogue pairs. There is a small effect in the low-middle mass range, but not so much at the high mass end. This is because orbits are more radial at high mass as there is less likely to be a third massive object that imposed a significant tidal torque. Because of this, adding the $\widetilde{b}$ condition makes the mass function flatter. Extrapolating this to higher masses can lead to nonsensical results such as obtaining a lower $\chi$ value \emph{after} imposing the $\widetilde{b} \leq \widetilde{b}_{EG}$ condition. Thus, we do not show $\chi$ values for the case of imposing both $\widetilde{v} \geq \widetilde{v}_{EG}$ and $\widetilde{b} \leq \widetilde{b}_{EG}$, except to note that the latter would not by itself make matters much worse for $\Lambda$CDM because radial orbits are expected at the high mass of El Gordo. We also do not consider results for the $\widetilde{v} > 0$ case to be particularly useful because requiring $\widetilde{v} > 0$ is not sufficient for finding analogues to a cluster merger that clearly requires a high infall speed to reproduce the observed shock features (Section~\ref{Infall_velocity_condition}). Similarly, imposing no condition on $\widetilde{v}$ is not sufficient to match El Gordo $-$ this is shown for illustrative purposes only.

\begin{figure}
	\centering
	\includegraphics[width = 8.5cm] {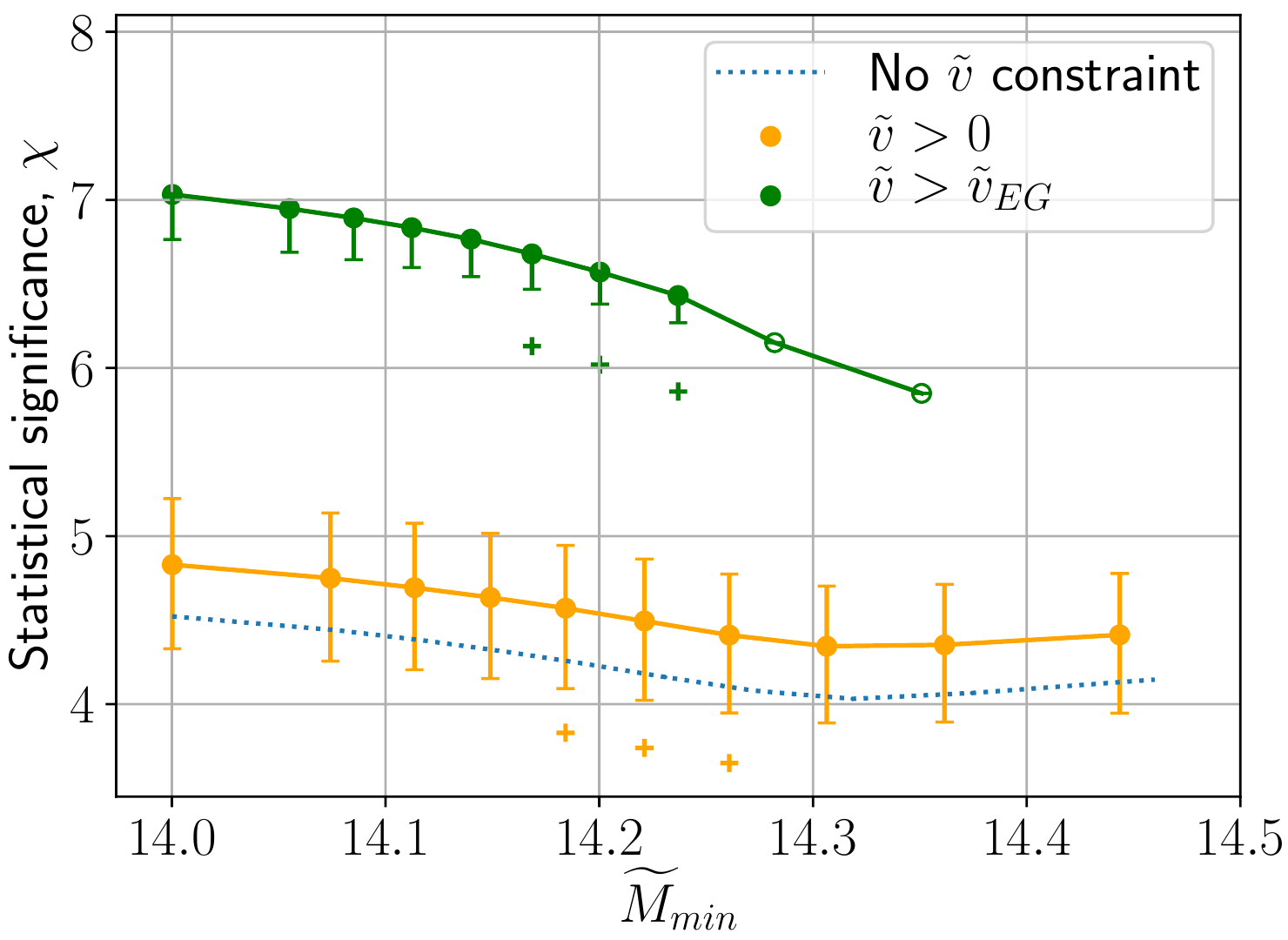}
	\caption{The statistical significance $\chi$ (Equation \ref{P-chi}) of the observation of an El Gordo-like object as defined in different ways, in terms of the minimum mass beyond which we fit the cumulative $\widetilde{M}$ distribution. Each line corresponds to a different set of conditions, which in increasing order of restrictiveness are: turnaround and mass ratio (blue dotted), $\widetilde{v} > 0$ (orange), and $\widetilde{v} > \widetilde{v}_{EG}$ (green). Error bars on the $\widetilde{v} > 0$ line show the effect of varying the El Gordo mass by $\pm20\%$, while those on the $\widetilde{v} > \widetilde{v}_{EG}$ line correspond to a $\pm10\%$ uncertainty on $\widetilde{v}_{EG}$. Along each line, each successive data point is based on 10\% fewer pairs at $z = 1$ in terms of the left-most point, e.g. if this is based on 100 pairs, the second is based on 90 pairs and the third on 80 pairs etc. The hollow circles indicate somewhat questionable results $-$ error bars could not be reliably calculated as there were too few pairs. This is also the reason for missing upper error bars on the line for $\widetilde{v} > \widetilde{v}_{EG}$. The cross-shaped symbols represent the results of our lightcone tomography (Section~\ref{Lightcone_tomography}), with the same relation between colour and the imposed conditions.} 
	\label{chi-mass}
\end{figure}

Our quadratic fits to the mass function may not be perfectly accurate as there could be a weak cubic dependence. Since the mass function declines rather steeply, our fits to it are dominated by pairs near the low mass end, causing the fits to prioritize very modest improvements here over a better fit at the high mass end critical to our analysis. One way to check for the impact of any cubic or higher order trend is to restrict the range of $\widetilde{M}$ used in our fit. By focusing on the high mass region, it is possible to minimise such systematic effects at the cost of higher random errors from Poisson noise. Therefore, we try several different values for $\widetilde{M}_{min}$, the lowest value of $\widetilde{M}$ used in our quadratic fit. The idea is to check if our results differ much when fitting the whole data set or fitting towards the high mass end. We have summarized the results of all these different tests in Figure~\ref{chi-mass}, where we have expressed the number of analogues expected to be found in the El Gordo survey volume in terms of the $\chi$ value. This does not depend very much on $\widetilde{M}_{min}$.

\subsection{Lightcone tomography}
\label{Lightcone_tomography}

To check that the previous analysis provides an accurate estimate of the probability that $\Lambda$CDM yields an El Gordo analogue in the surveyed volume, we compare the results with a different type of statistical analysis which we term lightcone tomography. The basic idea is to consider the distribution of pairs with El Gordo-like properties along our entire past lightcone, not just at $a \approx 0.5$. In particular, after requiring that pairs have similar dimensionless parameters to El Gordo, we consider to what extent its main dimensionful numbers (mass and redshift) are outliers to the distribution expected in $\Lambda$CDM.

The procedure can be described as follows: we apply our usual quadratic fit to the mass function (Equation \ref{Mass_function_fit}) in the Jubilee snapshots at $z = 0$, $z = 0.509$, and $z = 1$. The fit coefficients for the $\widetilde{v} > \widetilde{v}_{EG}$ condition are shown in Appendix~\ref{Appendix}. Since we have three snapshots, we use a quadratic to fit the values for each coefficient against $\ln a$, allowing us to extrapolate the value of this coefficient to any other $a$. In other words, $c_0$ in Equation \ref{Mass_function_fit} is now treated as a function of $a$, with its value $c_0 \left( a \right)$ found by a quadratic interpolation from the three snapshots where we know $c_0$. This tells us the co-moving number density of El Gordo-like pairs as a function of both $\widetilde{M}$ and $a$. For each bin in $a$ along our past lightcone, the co-moving volume is given by integrating Equation \ref{V_eff}.

\begin{figure}
	\centering
	\includegraphics[width = 8.5cm] {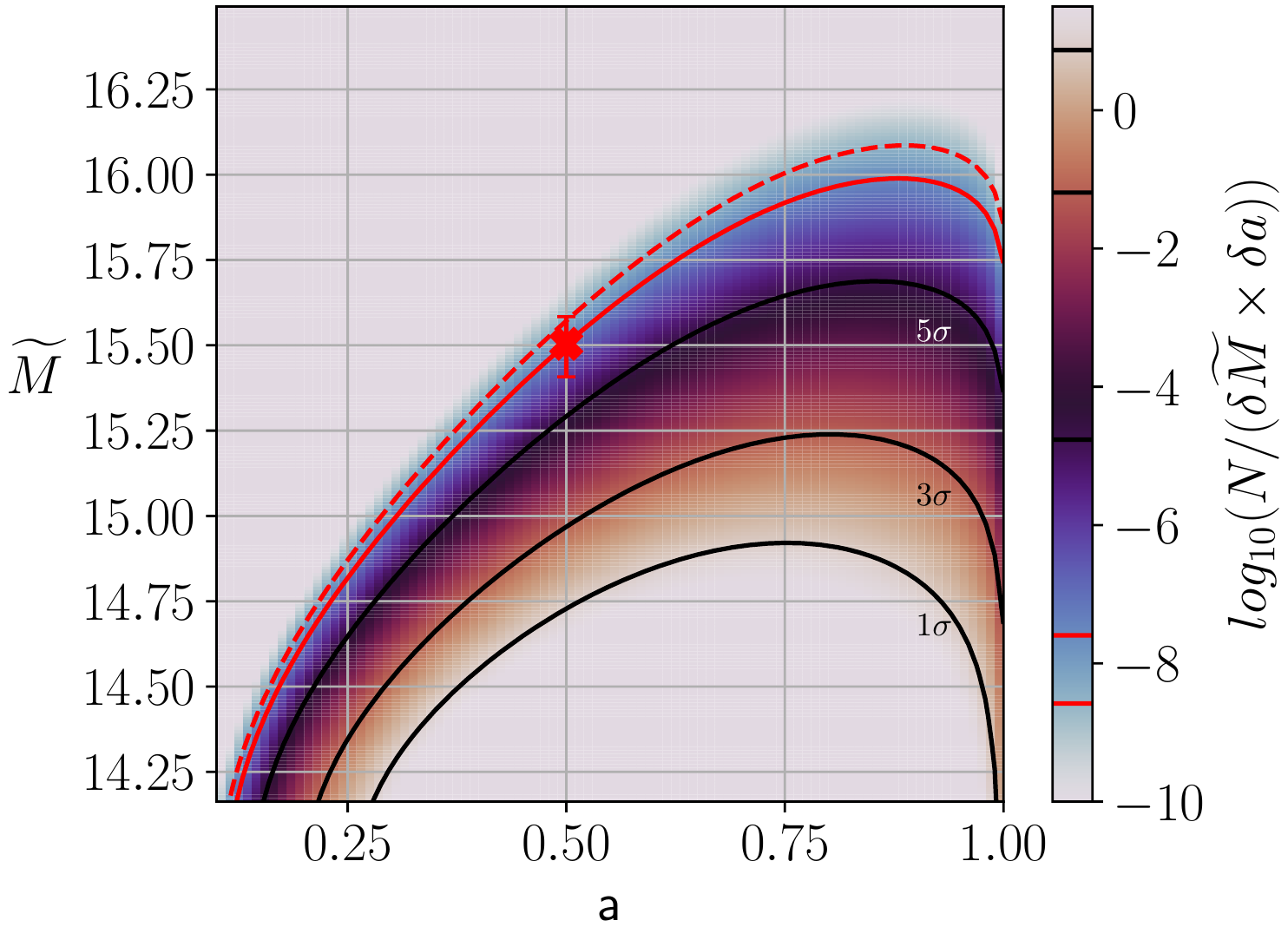}
	\caption{The number of analogues to El Gordo according to the turnaround, mass ratio, and $\widetilde{v} \geq \widetilde{v}_{EG}$ conditions per unit $\widetilde{M}$ per unit $a$ along our past lightcone for the 755~$deg^2$ sky area in which it was discovered. The black contour lines show the $1\sigma$, $3\sigma$, and $5\sigma$ confidence regions, while the red contour at $6.16\sigma$ goes through El Gordo (red cross, with error bar indicating $\pm 20\%$ uncertainty on the mass). Of the $N_{EG}$ pairs outside this contour, only $0.1 \, N_{EG}$ pairs lie beyond the red dashed contour, i.e. most of the probability of observing El Gordo comes from between the red contours. The levels of all these contours are shown on the colour bar. The $5\sigma$ contour is at $\widetilde{M} = 15.293$ when $a = 0.5$.} 
	\label{cont_vEG}
\end{figure}

We use this information to obtain the number of pairs in the past lightcone of a $z = 0$ observer that lie within any given range of pair total mass and redshift. Using a 2D contour plot showing this quantity as a function of $\left( \widetilde{M}, a \right)$, we get the contour level for El Gordo. We then infer the expected number $N_{EG}$ of El Gordo-like objects in our past lightcone with $\left( \widetilde{M}, a \right)$ outside this contour. This lets us infer the corresponding $P$ and $\chi$ values. Figure~\ref{cont_vEG} shows this contour plot for the $\widetilde{v} > \widetilde{v}_{EG}$ condition using a mass function fit with $\widetilde{M}_{min} = 14.16$. The solid red line shows the critical contour through El Gordo, beyond which we expect $N_{EG} = 7.51 \times 10^{-10}$ pairs. For ease of reference, we call this the $N_{EG}$ contour.

As El Gordo is clearly a significant outlier, most of these pairs lie only a little outside the $N_{EG}$ contour. This can be seen with the red dashed line, which shows the contour beyond which there are $0.1 \, N_{EG}$ pairs. Thus, fully 90\% of $N_{EG}$ arises from the thin strip between the red contours. Since the number density of pairs with respect to $\left( \widetilde{M}, a \right)$ is the same along each contour, the total number of pairs contributed by each $a$ is roughly proportional to the gap between the $N_{EG}$ and $0.1 \, N_{EG}$ contours $-$ the wider the gap, the more important that $a$. Thus, we can gain an idea of which cosmological epochs are most relevant to the probability of observing El Gordo in a $\Lambda$CDM universe. Figure~\ref{cont_vEG} shows that the gap between the $0.1 \, N_{EG}$ and $N_{EG}$ contours is widest for $a \ga 0.5$, which is also the range covered by the Jubilee snapshots we analyse. Therefore, even if our extrapolation to lower $a$ is not perfectly accurate, this should not significantly affect our results $-$ the total probability receives little contribution from $a \la 0.5$.

Our lightcone tomography analysis indicates that when requiring $\widetilde{v} > \widetilde{v}_{EG}$, the statistical significance $\chi = 6.16$. This agrees fairly well with our power-law analysis (Figure~\ref{chi-mass}). Importantly, \emph{both techniques agree that the $\Lambda$CDM model is falsified at $>6\sigma$ significance based on El Gordo alone.}

For our lightcone tomography, we consider cluster pairs all the way up to $a = 1$ in order to be conservative. This is not very realistic as there must have been some time lag between the observed state of El Gordo and the pre-merger stage at which we seek to capture El Gordo analogues in the Jubilee simulation. As explained in Section~\ref{Redshift_condition}, we estimate that this time lag is 559~Myr. Subtracting this from the 13.47~Gyr age of the universe in $\Lambda$CDM yields $a = 0.96$. In the real world, any El Gordo-like pair at $a > 0.96$ would simply not have enough time to evolve into the observed state. Thus, it may be more reliable to restrict our lightcone tomography to $a < 0.96$. This would slightly reduce $N_{EG}$ and increase the significance $\chi$. We found that $\chi$ rises by only $\approx 0.01$, so the time lag effect is not a major source of uncertainty at the present epoch. It may be more significant at $a \approx 0.5$, but we have already accounted for it by placing El Gordo at $z = 1$ instead of the observed $z = 0.87$ (Section~\ref{Redshift_condition}). As discussed there, the time lag could be much longer than our assumed 560~Myr since most of the candidate pairs we identify have a rather large separation, so they would need a significant amount of time to reach pericentre (Figure~\ref{hist_dist}). Moreover, the time after pericentre could by itself contribute 910 Myr if El Gordo is observed after second turnaround, which better accounts for some observables \citep{Karen_2015}.

\subsection{Combined implications with the Bullet Cluster}
\label{Combined_with_BC}

\renewcommand{\arraystretch}{1.2}
\begin{table*}
	\centering
	\begin{tabular}{c|c|c|c|c|}
		\cline{2-5}
		& \multicolumn{2}{c|}{$\widetilde{v}>0$}            & \multicolumn{2}{c|}{$\widetilde{v} > \widetilde{v}_{EG}$} \\ \cline{2-5} 
		& Power-law     & Lightcone            & Power-law     & Lightcone           \\
		&               & tomography           &               & tomography          \\ \hline
		\multicolumn{1}{|c|}{$P_{EG}$}                    & $4.08 \times 10^{-6}$ ($4.61\sigma$) & $1.12 \times 10^{-4}$ ($3.86\sigma$) & $2.23 \times 10^{-11}$ ($6.69\sigma$) & $7.51 \times 10^{-10}$ ($6.16\sigma$)  \\ \hline
		\multicolumn{1}{|c|}{$P_{EG+BC}$}                 & $5.13 \times 10^{-7}$ ($5.02\sigma$) & $1.20 \times 10^{-5}$ ($4.38\sigma$) & $3.98 \times 10^{-12}$ ($6.94\sigma$) & $1.24 \times 10^{-10}$ ($6.43\sigma$) \\ \hline
		\multicolumn{1}{|c|}{$P_{EG+BC}$ (full sky)}      & $6.09 \times 10^{-5}$ ($4.01\sigma$) & $1.36 \times 10^{-3}$ ($3.20\sigma$) & $5.04 \times 10^{-10}$ ($6.22\sigma$) & $1.55 \times 10^{-8}$ ($5.66\sigma$) \\ \hline
	\end{tabular}
	\caption{Results of the power-law (Section~\ref{Power_law_method}) and lightcone tomography (Section~\ref{Lightcone_tomography}) analyses for different $\widetilde{v}$ conditions. We show the expected number of analogues $N$ in the surveyed region as a $P$-value $\left[ P = 1 - \exp \left(-N \right) \right]$ and the corresponding statistical significance $\chi$ (Equation \ref{P-chi}). Results for El Gordo (the Bullet Cluster) have a subscript $_{EG}$ ($_{BC}$). The row $P_{EG+BC}$ shows the effect of combining our results with the Bullet Cluster, which is a $\chi_{_{BC}} = 2.78\sigma$ outlier according to the results of \citet{Kraljic_2015} when accounting for the sky footprint of the discovery survey (see text). We also show the impact of scaling the number of El Gordo and Bullet Cluster analogues by assuming that the surveys in which they were found covered the full sky (last row). This is a conservative estimate of how our conclusions would be weakened if the entire sky contains no additional problematic objects for $\Lambda$CDM.}
	\label{Results_table}
\end{table*}
\renewcommand{\arraystretch}{1}

The Bullet Cluster was discovered by \citet{Tucker_1995} and is listed in their table~1. In their section~3, they indicate that this and the other clusters they discovered were identified by searching through 1435 fields of view of the Imaging Proportional Counter on the Einstein X-ray observatory \citep{Giacconi_1979}. According to table~1 of the latter work, each field of view is a square with $1.25^{\circ}$ sides. Thus, the Bullet Cluster was discovered in only 5.4\% of the sky, with the small sky coverage being due to poor spatial resolution that meant much of the sky was obscured by emission from foreground supernova remnants. Despite these difficulties, the extreme properties of the Bullet Cluster were already apparent in \citet{Tucker_1998}.

A detailed analysis of the Bullet Cluster is beyond the scope of this contribution, but we utilize the result of \citet{Kraljic_2015} that 0.1 pairs analogous to it should be observed out to its redshift of $z = 0.296$. They implicitly assumed that observing a cluster pair `as bad as or worse than' the Bullet involved a lower redshift, which is possible for a nearby object since the co-moving volume per unit $a$ declines rapidly with $a$ when $a \approx 1$. We have made the opposite assumption since El Gordo is at a much higher redshift, so it is clear that the situation will be worsened for $\Lambda$CDM by going to higher $z$. The situation may be different for the Bullet, so we assume that the choice adopted by \citet{Kraljic_2015} is reasonable. However, their result of 0.1 analogous pairs is valid only for a survey that covers the full sky. This is apparent in their equation 4.5, which is directly analogous to our Equation \ref{V_eff} $-$ but with $4\mathrm{\pi}$ instead of the sky area $A$. Accounting for the sky coverage of 5.4\%, we get that the surveyed region is expected to have $N_{BC} = 5.4 \times 10^{-3}$ pairs analogous to the Bullet Cluster in a $\Lambda$CDM context. This makes the Bullet Cluster a $\chi_{_{BC}} = 2.78\sigma$ outlier.\footnote{This result could differ somewhat if \citet{Kraljic_2015} had used the dimensionless $\widetilde{v}$ instead of the infall velocity, since cluster pairs of lower mass (which are more common) can more easily match $\widetilde{v}$ than the actual infall velocity.}

To approximately combine this with El Gordo, we add the squares of the individual $\chi$ values.
\begin{eqnarray}
	\chi^2_{tot} ~=~ \chi_{_{BC}}^2 + \chi_{_{EG}}^2 \, .
	\label{Combining_chi_sq}
\end{eqnarray}
The probability of a higher $\chi^2_{tot}$ is then found using the standard formula for two degrees of freedom.
\begin{eqnarray}
	P ~=~ \exp \left( -\frac{\chi^2_{tot}}{2} \right) \, .
	\label{chi_sq_2dof}
\end{eqnarray}
This can be converted into a statistical significance $\chi$ for a single variable using our usual approach of applying Equation \ref{P-chi}.

Table~\ref{Results_table} summarizes the main results of the power-law and lightcone tomography analyses, both when considering El Gordo alone and in combination with the Bullet Cluster. We see that the power-law and lightcone tomography analyses give similar results. The Bullet Cluster and El Gordo were discovered in a small fraction of the entire sky. To be conservative, we can assume that no additional cluster pairs are found in the rest of the sky that pose a problem to $\Lambda$CDM. We illustrate what effect this could have by scaling up $N_{BC}$ and $N_{EG}$ under the assumption that their discovery surveys covered the whole sky. This reduces $\chi_{_{BC}}$ and $\chi_{_{EG}}$, but we are still left with an overall significance $\chi > 5$. Therefore, the power-law and lightcone tomography methods agree that \emph{the $\Lambda$CDM model must be rejected at $>5\sigma$ confidence even if no additional problematic objects exist in the rest of the sky.}

In reality, full sky surveys such as the Planck survey \citep{Planck_VIII_2011} have already found more clusters that could further increase the tension. One example is PLCK G214.6+37.0, a triple system of clusters at $z \approx 0.47$ that appear to be at an early stage of interaction. The total mass $M_{200} \approx 1.17 \times 10^{15} M_{\odot}$, while the mass ratio between its components is $\leq 1.4$. Another example is PLCK G287.0+32.9 at $z = 0.39$ \citep{Planck_VI_2013}, a quadruple system of clusters with total mass $M_{200} = 2.04^{+0.20}_{-0.21} \times 10^{15} M_{\odot}$ in which the largest cluster ($10\times$ more massive than the other components of the system) is undergoing a complex merger with one or more of the other clusters \citep{Bagchi_2011, Finner_2017}. We discussed a few other potentially problematic cases in Section~\ref{Introduction}. It is therefore unrealistic to assume that there are no other objects besides El Gordo and the Bullet Cluster that could entail a problem for $\Lambda$CDM. Moreover, even this very conservative assumption is insufficient to solve the tension (Table~\ref{Results_table}). 

\section{Possible explanations for El Gordo}
\label{Discussion}

Table~\ref{Results_table} shows that the results from our power-law and lightcone tomography analyses are in agreement, with a difference in statistical significance $\chi$ of $\approx 0.7$ when requiring $\widetilde{v} > 0$ and $\approx 0.5$ for $\widetilde{v} > \widetilde{v}_{EG}$. Our most conservative estimate of the tension is $\chi_{_{EG}} = 6.16$ based on lightcone tomography, so an uncertainty of 0.5 would not be enough to reconcile $\Lambda$CDM with observations at $5\sigma$ confidence.\footnote{If we assign our calculated $\chi$ a Gaussian uncertainty of 0.5, then we can suppose that we overestimate $\chi$ by $0.5y$. The overall likelihood of any $y$ is then that of $\chi^2_{tot} = y^2 + \left(6.16 - 0.5 y \right)^2$ for two degrees of freedom. We never get $\chi^2_{tot} < 30.4$, but a $5\sigma$ event corresponds to $\chi^2_{tot} = 28.7$ (Equation \ref{chi_sq_2dof}).} Moreover, if $\chi_{_{EG}}$ is combined with the $2.78\sigma$ tension caused by the Bullet Cluster, the combined significance rises to $6.43\sigma$. These results account for the sky coverage of the respective surveys in which the objects were discovered. If we adjust our analysis by pretending the surveyed area in each case was the whole sky, the significance is still close to $6\sigma$. Thus, future surveys will not be able to alleviate the tension for $\Lambda$CDM even if they uncover no additional problematic objects.

Even with the more relaxed condition ($\widetilde{v} > 0$) in which the high infall velocity of El Gordo is not taken into account, we still get a statistical significance close to $4\sigma$, which makes the El Gordo mass at $z = 1$ an extremely unlikely feature on its own. Isolated clusters of this mass are not catastrophic for $\Lambda$CDM \citep[e.g. figure~17 of][]{Jee_2014}. However, the paired nature of El Gordo substantially reduces the odds of finding such a system in the surveyed region. This very important effect was not considered in previous works that treated El Gordo as one object when quantifying the tension with $\Lambda$CDM \citep{Menanteau_2012, Jee_2014, Sahlen_2016}. Of course, these analyses could not consider the high required infall velocity. Our analysis is the first to consider both the $\widetilde{M}$ and $\widetilde{v}$ of El Gordo in a standard context.

Even if El Gordo is treated as one object, the statistical methods used differ between studies. Unlike \citet{Jee_2014}, \citet{Buddendiek_2015} showed in their figure~8 that even a single $2 \times 10^{15} M_\odot$ cluster at $a = 0.5$ would be quite problematic for $\Lambda$CDM. This is also evident in figure~5 of \citet{Watson_2014}. Regardless of the exact frequency of such clusters at this redshift, a major merger of two such clusters would be even less frequent, since this can only occur if two individually rare objects formed close enough to turn around from the cosmic expansion within a few Gyr. It is thus not very surprising that the existence of El Gordo significantly challenges $\Lambda$CDM.

Beyond this qualitative agreement with previous studies, we are unable to compare our results with other similar works since they do not conduct an analysis nearly as detailed as ours. However, we can compare the results of our two statistical analyses among themselves. Both give $\chi > 6$, so we can be fairly confident in setting the lower limit of the $\chi$ value to 6. Therefore, the expected number of analogues to El Gordo in the surveyed region must be increased by $>3$ orders of magnitude just to reach the plausibility threshold of $\chi = 5$, which would make the observation of El Gordo as probable as a fair coin landing heads 21 times in a row.

The rather low $P$-values we obtain are almost certainly overestimates for various reasons. As discussed in Section~\ref{Redshift_condition}, many of the Jubilee pairs we call `analogues' are actually still rather widely separated at $z = 1$, so they would not evolve into the observed configuration by $z = 0.87$. Moreover, we have only searched for pairs whose orbit is similar to that of the presumed El Gordo progenitors. We have not allowed for the fact that only a small fraction of the orbit is spent close to pericentre, but we must be observing El Gordo at this phase. The $P$-values are also reduced by the requirement to have a sufficiently small impact parameter, though this by itself is not very problematic (Figure~\ref{allcond}).

\subsection{Reliability of the Jubilee simulation}
\label{Jubilee_reliability}

The length and mass scales relevant to El Gordo should be very well resolved in the Jubilee simulation: the minimum resolved halo mass is $1.49 \times 10^{12} \, h^{-1} M_{\odot}$ \citep[20 particles; see section~2 of][]{Watson_2014b}, and the minimum spatial resolution is $0.5 \, h^{-1}$~cMpc \citep[mesh number of $12000^3$ for a simulation box with side length of $6 \, h^{-1}$~cGpc; see table~1 of][]{Watson_2013}. At $z = 1$, these correspond to a virial mass and radius of $M_{200} = 1.88 \times 10^{12} M_{\odot}$ and $r_{200} = 0.3$~Mpc, respectively. Since we select only those haloes which have $M_{200} \geq 3.5\times 10^{13} M_{\odot}$, we can be confident that the haloes we use are all well resolved.

The 8.57~cGpc side length of the Jubilee simulation significantly exceeds the co-moving Hubble radius $c/\left( aH \right) = 5.04$~cGpc at $a = 0.5$ ($c$ is the speed of light in vacuum), so the simulation is about as large as the entire observable Universe at that epoch. As a result, the Jubilee simulation volume is $\approx 1380 \times$ larger than that of the El Gordo discovery survey. Therefore, the severe tension between $\Lambda$CDM and the existence of El Gordo is evident already in the disparity between its mass and that of the most massive Jubilee pair with similar dimensionless parameters (Figure~\ref{allcond}). Importantly for the accuracy of our results, they are subject to only a very small amount of cosmic variance due to the large simulation volume $-$ if the root mean square (rms) matter density fluctuation between spheres of radius $8 \, h^{-1}$~Mpc is $\sigma_8 = 0.811 \pm 0.006$ today \citep{Planck_VI_2020}, then for the Jubilee volume at $z = 1$, it should be only $5.4 \times 10^{-4}$ as the density fluctuations scale inversely with size \citep{Harrison_1970, Zeldovich_1972} and grow $\appropto a$. Consequently, the density fluctuations on a mass scale of $10^{15} M_\odot$ are accurately handled by a numerical simulation with side length of just $0.1 \, h^{-1}$~cGpc \citep[figure~2 of][]{Watson_2013}.

Our own statistical analysis at $z = 1$ (from which we obtain our nominal result for $P_{EG}$) is based on fitting to 15035 pairs that pass the $\widetilde{v} \geq \widetilde{v}_{EG}$ condition and whose total mass $\widetilde{M} \geq 14.16$ to ensure only clusters are considered. The highest mass pair has $\widetilde{M} = 14.91$. Due to the large number of objects, the Poisson noise should be very small over the majority of this range, allowing for an accurate analytic fit to the pairwise mass function. This is evident from the fractional uncertainty in the number of pairs with $\widetilde{M} \geq 14.16$: our estimate of $N = 15035$ in the simulation volume has a fractional Poisson uncertainty of only
\begin{eqnarray}
	\textrm{Poisson noise} ~=~ \frac{1}{\sqrt{N}} ~=~ 8.16 \times 10^{-3} \, .
	\label{Pnoise}
\end{eqnarray}
This explains the very good analytic fit in Figure~\ref{allcond}.

The uncertainty increases at higher mass $-$ due to the extreme rarity of objects like El Gordo in Jubilee, it is not possible to simply count the number of analogue pairs with $\widetilde{M} > \widetilde{M}_{EG}$. Instead, we must rely on an extrapolation of our quadratic mass function. Since we identified low mass analogues to El Gordo with $\widetilde{M}$ as high as 14.91, we only need to extrapolate the mass function by 0.6~dex, less than the range of $\widetilde{M}$ used to define the mass function. Thus, the extrapolated number of analogues with $\widetilde{M} > \widetilde{M}_{EG}$ should be quite reliable. The uncertainties would be even lower when $z = 0.509$ and $z = 0$ due to an increasing number of haloes above a fixed mass (Figure~\ref{cont_vEG}; see also Appendix \ref{Appendix}). 

The majority of the pairs we find are separated by much more than the sum of their virial radii (Figure~\ref{hist_dist}). As a result, the halo finder used (\textsc{ahf}) would only very rarely mis-classify a pair as one object, incorrectly removing it from the statistics. Different halo finders also yield similar cluster mass functions \citep[figure~3 of][]{Watson_2014}. Therefore, we can be confident that simulated galaxy cluster pairs similar to the El Gordo progenitors are reliably identified by \textsc{ahf}.

For these reasons, we expect that the pairwise mass function of the Jubilee simulation over the fitted range should be a very accurate representation of $\Lambda$CDM. Indeed, \citet{Watson_2014} noted that the Jubilee simulation has proven capable of reproducing the distribution of massive individual clusters according to both observations and theoretical predictions. It has also complied with most of the $\Lambda$CDM predictions for the distribution of structures.

Currently, it would be hard to compare if other cosmological simulations give a similar occurrence rate for extreme objects like El Gordo, since cosmological simulations as large as Jubilee are not that common. Nevertheless, there is no reason to believe that the Jubilee simulation might not accurately represent $\Lambda$CDM since, up to now, it has been shown to work correctly in accordance with the $\Lambda$CDM cosmological model for which it was designed. Moreover, the use of Poisson statistics is justified for rare objects like El Gordo \citep{Watson_2014}.

\subsection{A lower mass}
\label{Lower_mass}

Our adopted mass of $3.2 \times 10^{15} M_\odot$ exceeds the $(2.16 \pm 0.32) \times 10^{15} M_{\odot}$ estimated in table~2 of \citet{Menanteau_2012} by combining various techniques. However, the large uncertainties given there indicate that these earlier estimates are not very reliable. In addition, the methods used to obtain the El Gordo mass did not account for the fact that it is an interacting cluster. The X-ray temperature and luminosity-based estimates of $\approx 2.6 \times 10^{15} M_\odot$ should have been multiplied by a factor of 1.17 to account for this \citep[section~4.1.3 of][]{Vikhlinin_2009}. In addition, the SZ signal mass estimate of $1.6^{+0.6}_{-0.4} \times 10^{15} M_\odot$ is biased low by $\approx 10-20\%$ in the case of interacting clusters \citep[figure~7 of][]{Krause_2012}, while the estimate from the cluster's velocity dispersion should have taken into account that dynamical friction slows down a substructure before it can suffer mass loss due to tidal stripping \citep{Munari_2013}. In fact, \citet{Menanteau_2012} mention in their section~3.3 that due to considering El Gordo a virialized cluster, their nominal value for its mass is likely to be an underestimate. This is supported by the fact that summing their dynamical masses for its two subcomponents gives $(2.8 \pm 0.9) \times 10^{15} M_{\odot}$, which is closer to the \citet{Jee_2014} weak lensing estimate. Clearly, reliable estimates of the El Gordo mass require a hydrodynamical model designed to reproduce the X-ray, weak lensing, and SZ maps in detail. This is precisely what was done by \citet{Zhang_2015}, as discussed in their section~2.2.

So far, we have focused on their fiducial Model B for reasons discussed in Section~\ref{Models}. Figure~\ref{cumul_mass} shows that the lower mass Model A may reduce the tension with $\Lambda$CDM. We therefore discuss whether it provides a better explanation of the observations in a $\Lambda$CDM context. This entails considering the likelihood of the pre-merger configuration and how well Model A reproduces the observed morphology.

\begin{figure}
	\centering
	\includegraphics[width = 8.5cm] {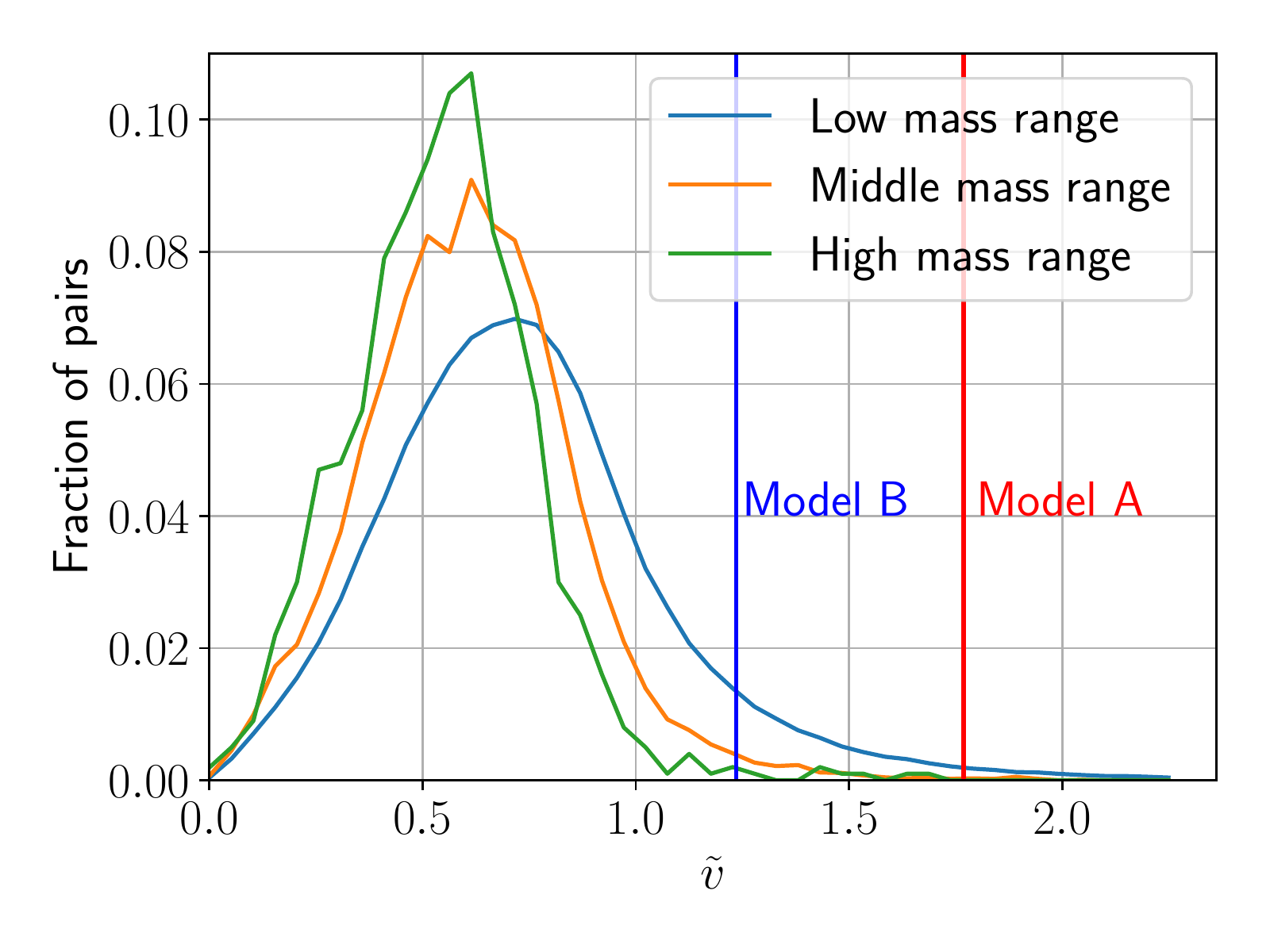}
	\caption{The $\widetilde{v}$ distribution for different mass ranges (in units of $10^{15} M_\odot$: $1-3.6$, $3.6-6.2$, and $>6.2$). The `low' and `middle' mass ranges each have 396350 pairs, while the `high' mass range has only 1000 pairs. All curves are normalized to a sum of 1. The vertical lines show the $\widetilde{v}$ of each model. Notice the narrowing of the $\widetilde{v}$ distribution with pair total mass.}
	\label{vel_distr}
\end{figure}

Although Model A requires a lower mass for El Gordo ($\widetilde{M} = 15.29$), the progenitors should have a more nearly equal mass ratio ($\leq 2$ rather than $\leq 3.6$) and be infalling onto each other $\approx 20\%$ faster \citep[table~2 of][]{Zhang_2015}. Thus, Model A requires $\widetilde{v} = 1.77$ while model B manages with a lower $\widetilde{v} = 1.24$. To explore what effect this could have, we use Figure~\ref{vel_distr} to show the distribution of $\widetilde{v}$ in three different $\widetilde{M}$ bins. It is clear that the higher $\widetilde{v}$ will significantly reduce the number of analogues if the Model A parameters are adopted. This effect is especially pronounced at the high mass of El Gordo because the $\widetilde{v}$ distribution becomes narrower at high masses. The most likely explanation is that such massive cluster pairs are probably quite isolated $-$ it would be very unlikely to find a third massive object in the vicinity of the pair. Without the third object, it is very difficult for the pair to have $\widetilde{v} > 1$ as this entails more kinetic energy than is required to escape. This is why the observed combination of $\widetilde{M}$ and $\widetilde{v}$ is so problematic for $\Lambda$CDM.



\begin{figure}
	\centering
	\includegraphics[width = 8.5cm] {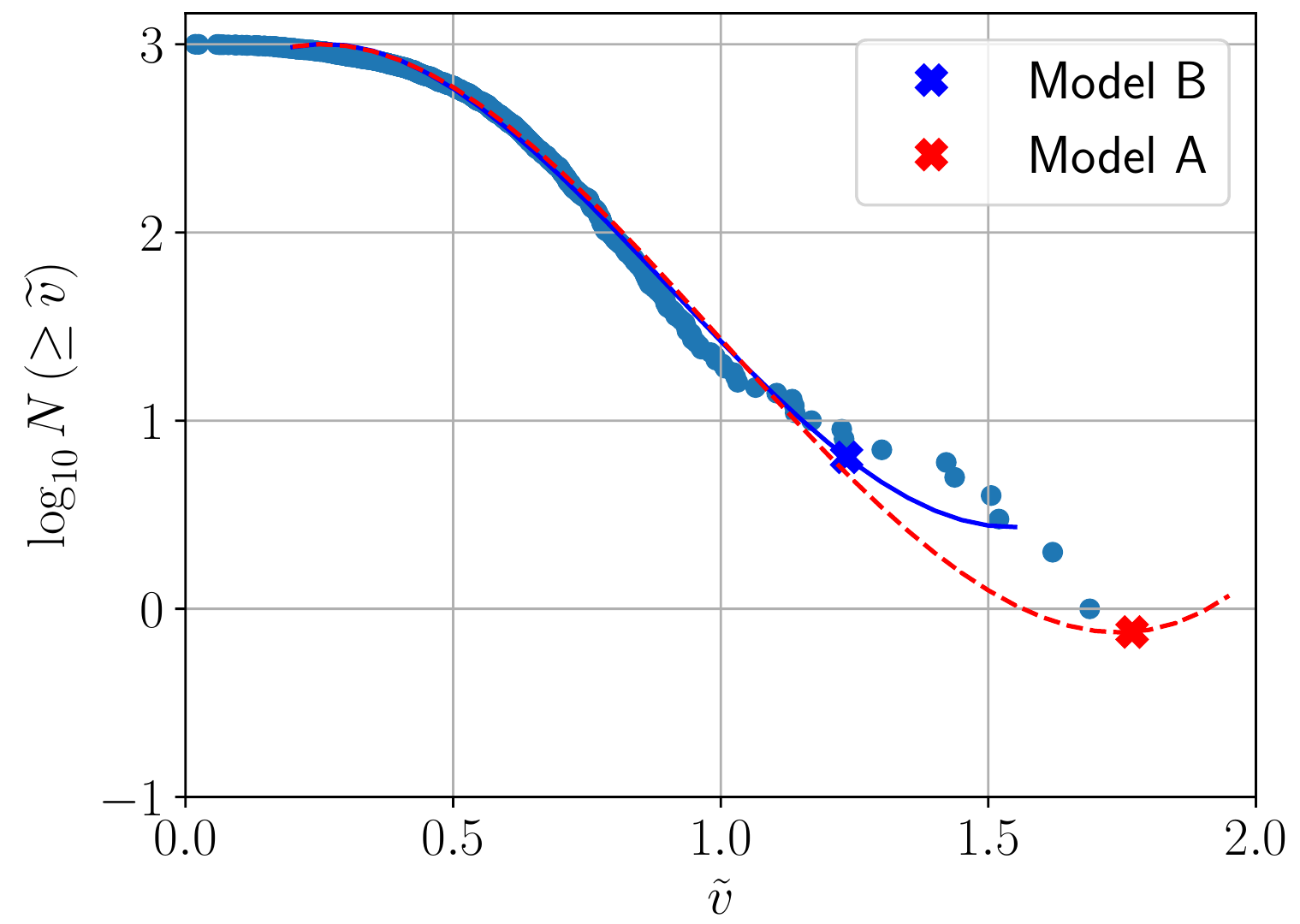}
	\caption{The cumulative $\widetilde{v}$ distribution for the 1000 most massive candidate El Gordo analogues. The dotted red (solid blue) line shows a cubic fit for Model A (B). For clarity, individual points are only shown for the more realistic Model B (see text). The crosses show $\widetilde{v}$ for each model. Notice that Model A is more of an outlier here, opposite to the situation in Figure~\ref{cumul_mass}.}
	\label{cumul_vel} 
\end{figure}

Model A thus achieves a different trade-off between how difficult it is for $\Lambda$CDM to explain different aspects of the pre-merger configuration. To better understand if it helps overall, we fit the cumulative $\widetilde{v}$ distribution $\log_{10} N \left( \geq \widetilde{v} \right)$ using a cubic (Figure~\ref{cumul_vel}). It is evident that the higher $\widetilde{v}$ reduces the number of $\widetilde{v}$ analogues by $\approx 1$ order of magnitude, as also noted in Table~\ref{analogs_mv}. This is less significant than the $\approx 2$ orders of magnitude increase expected from the use of a lower mass (Figure~\ref{cumul_mass}), suggesting that Model A is more likely from a cosmological perspective.

\renewcommand{\arraystretch}{1.1}
\begin{table}
	\centering
	\begin{tabular}{c|c|c|c|}
		\cline{2-4}
		& \multicolumn{3}{c|}{Number of pairs with higher} \\
		& $\widetilde{M}$ & $\widetilde{v}$ & Both \\ \hline
		\multicolumn{1}{|c|}{Model A} & 1.23 & $7.53 \times 10^{-1}$ & $1.00 \times 10^{-5}$ \\
		\multicolumn{1}{|c|}{Model B} & $1.42 \times 10^{-2}$ & 6.38 & $3.08 \times 10^{-8}$ \\ \hline
	\end{tabular}
	\caption{The number of pairs with larger $\widetilde{M}$ and/or $\widetilde{v}$ than El Gordo in the Jubilee simulation volume of $\left( 6 \, h^{-1} \, \text{cGpc} \right)^3$, which is much larger than the surveyed region (Section~\ref{Power_law_method}). The properties of El Gordo are from Models A and B in \citet{Zhang_2015}. Since the $\widetilde{v}$ distribution narrows at high masses, the results for $\widetilde{v}$ are based on cubic fits to the cumulative $\widetilde{v}$ distribution of the 1000 most massive pairs (Figure \ref{cumul_vel}). The last column corresponds to imposing the $\widetilde{v}$ condition before performing a quadratic extrapolation in $\widetilde{M}$. Since each condition is individually rather problematic, requiring both simultaneously leads to very few analogues (see text).}
	\label{analogs_mv}
\end{table}
\renewcommand{\arraystretch}{1}

However, the last column of Table~\ref{analogs_mv} shows that it is still very difficult to simultaneously explain the mass and collision velocity of El Gordo, even with the Model A parameters. The very low number of analogues listed there can be understood as follows: of the 1.23 analogues in $\widetilde{M}$ for Model A, we expect a fraction 0.753/1000 to pass the $\widetilde{v}>\widetilde{v}_{EG}$ condition, implying that $9.26 \times 10^{-4}$ systems pass both conditions if they are independent. However, the results in Figure~\ref{vel_distr} show that the $\widetilde{v}$ distribution gets narrower at high $\widetilde{M}$. We implicitly account for this narrowing by only considering pairs with $\widetilde{v}>\widetilde{v}_{EG}$ before extrapolating in $\widetilde{M}$. Thus, we expect even fewer pairs with both a higher $\widetilde{M}$ and a higher $\widetilde{v}$ than El Gordo. This probably explains why we get only $1.00 \times 10^{-5}$ analogues instead of $9.26 \times 10^{-4}$.

To quantify the overall effect of using Model A instead of Model B, we repeat our lightcone tomography analysis for the Model A parameters. This method suggests less tension with $\Lambda$CDM, so we use it to be more conservative. We find that the $P$-value is $2.73 \times 10^{-7}$, representing $5.14\sigma$ tension. The $360\times$ enhancement to the Model B probability of $7.51 \times 10^{-10}$ (Section \ref{Lightcone_tomography}) is very much in line with the $320\times$ enhancement suggested by the last column of Table \ref{analogs_mv}. When combined with the Bullet Cluster observations using Equation \ref{Combining_chi_sq}, we get an overall tension of $5.50\sigma$. Thus, it is clear that the more likely pre-merger configuration in Model A is still inconsistent with $\Lambda$CDM expectations.

So far, we have only discussed whether the Model A pre-merger configuration is likely to arise in $\Lambda$CDM. For a holistic discussion of whether Model A is a valid solution in this context, we also need to consider whether it reproduces observations of El Gordo (second step in Figure \ref{Logic_diagram}). The fact that it does not do so nearly as well as Model B explains why \citet{Zhang_2015} adopted the latter as their nominal model and optimized it further in Extended Model B. The main problem with Model A is that it does not reproduce the twin-tailed morphology of El Gordo, as can be seen in their figure~6. This is a very important feature of the real El Gordo \citep[figure~1 of][]{Menanteau_2012}. In addition, the implied mass of $1.95 \times 10^{15} M_\odot$ is in tension with the weak lensing mass of $\left( 3.13 \pm 0.56 \right) \times 10^{15} M_\odot$ \citep{Jee_2014}. This estimate is for a slightly different definition of the mass: the radius used is such that the enclosed density is $200\times$ the average matter density rather than the critical density. However, the Universe was matter-dominated at $z = 1$, so the difference is small. This is evident in figure~14 of \citet{Jee_2014}, which shows that the quoted uncertainty is significantly larger.

Returning to the Model B case, we need to consider whether a lower mass version of Model B might plausibly agree with detailed observations, even if a higher mass yields better agreement. Clearly, if $\Lambda$CDM is correct, a compromise is needed between the likelihood of the initial conditions of a hydrodynamical merger simulation and the extent to which it matches observations in detail (Figure~\ref{Logic_diagram}). We can address this by considering the variations tried by \citet{Zhang_2015}, exploiting the fact that they ran 123 models. At the end of their section~3.2, they discuss a modification with respect to fiducial Model B involving a reduced main cluster mass of $\widetilde{M}_1 = 15.20$ and, because of the smaller size of the system, a lower impact parameter of $b = 600$~kpc. The mass ratio and infall velocity remain the same as in fiducial Model B, so the total mass is $\widetilde{M} = 15.31$ and $\widetilde{v} = 1.43$. Since this model has an X-ray luminosity significantly smaller than that of fiducial Model B, we can consider $\widetilde{M} = 15.31$ a secure lower limit without some other compensatory adjustment. Repeating our lightcone tomography analysis for these `low-mass Model B' parameters, we obtain a $5.17\sigma$ tension ($P = 2.35 \times 10^{-7}$).

\citet{Zhang_2015} concluded that Model B mergers require a higher mass of $\widetilde{M}_1 \approx 15.40$ in order to reproduce an X-ray luminosity consistent with observations. The X-ray luminosity can also be increased by considering a higher velocity ($\approx 3500$~km/s), but this would simply lead to a configuration very similar to Model A $-$ which as mentioned above also does not have plausible initial conditions, and is problematic for other reasons. In any case, the higher resulting $\widetilde{v}$ would worsen the tension for $\Lambda$CDM. Furthermore, Model A and low-mass Model B require a smaller impact parameter than fiducial Model B. This is geometrically less likely and ought to reduce the probability further by a factor of $\approx \left(3/8\right)^2$ in the Model A case and $\approx \left(6/8\right)^2$ in the low-mass Model B case. Therefore, a lower mass does not alleviate the tension between $\Lambda$CDM and the observed properties of El Gordo. While the pre-merger configuration becomes somewhat more likely in terms of mass, several additional tensions emerge when switching to either model.

\citet{Zhang_2015} were already aware that the high mass and collision velocity of their best-fitting model would be problematic for $\Lambda$CDM (see their section~4). Given also their thorough exploration of parameter space, we assume that further hydrodynamical $\Lambda$CDM simulations with significantly different initial conditions would not reproduce the observed properties of El Gordo. In particular, the tension cannot be reduced below $5\sigma$ by lowering the El Gordo mass, either by switching to Model A or while other parameters are fixed to Model B values. This is because substantially different pre-merger configurations fail to reproduce key aspects of the observations.

Recent surveys reveal an increasing number of high-$z$ massive objects that together could pose a problem for $\Lambda$CDM \citep[e.g.][]{Kang_2015} in addition to individual objects which by themselves are already on the verge of falsifying it. Thus, several authors considered the $\Lambda$CDM model to be the actual problem, and tried to explain the presence of these objects using a different model: \citet{Angus_2008} and \citet{Katz_2013} in Milgromian dynamics \citep[MOND;][]{Milgrom_1983}, and \citet{Brownstein_2007} in Modified Gravity \citep[MOG;][]{Moffat_2006}. MOG is a covariant scalar-tensor-vector gravity theory which allows the gravitational constant $G$, a vector field coupling $\omega$, and the vector field mass $\mu$ to vary with space and time. This model was recently ruled out at $5.49 \sigma$ using the velocity dispersion profile of the ultra-diffuse galaxy Dragonfly 44 \citep{Haghi_2019}. While perhaps the basic ideas of MOG can still be saved by applying some corrections or modifications, its current formulation seems to be in conflict with observations. Therefore, we do not investigate if MOG could work as a solution to the presence of massive, high-$z$ colliding clusters.

\subsection{Milgromian dynamics (MOND)}
\label{MOND}

The MOND model generalizes gravity at low accelerations in such a way that, in its regime of action ($g \ll a_0$), the gravitational field strength $g$ behaves as
\begin{eqnarray}
	g ~=~ \frac{\sqrt{GMa_{_0}}}{r}
	\label{Deep_MOND_limit}
\end{eqnarray}
at distance $r$ from an isolated point mass $M$ generating the gravitational field. MOND adds a fundamental new acceleration scale $a_0$ below which the deviation from Newtonian dynamics becomes significant. To match observed galaxy rotation curves, $a_0 \approx 1.2 \times 10^{-10}$~m/s$^2$ \citep{Begeman_1991, Gentile_2011}. This generalization of gravity has a direct implication on the measurement of masses both by dynamical methods and by lensing, which works similarly to General Relativity in that both have the same relation between the non-relativistic $\bm{g}$ and light deflection \citep[][and references therein]{Milgrom_2013}. The Newtonian dynamical mass $M_{_N}$ is related to the MOND dynamical mass $M_{_M}$ by \citep{Katz_2013}:
\begin{eqnarray}
	M_{_M} ~=~ {M_{_N}}^2 \times \left( \frac{a_{_0}r^2}{G} + M_{_N} \right)^{-1} \, .
	\label{M_m-M_n}
\end{eqnarray}
This was derived by \citet{Angus_2011} based on the simple interpolating function \citep{Famaey_2005}. This states that in spherical symmetry, the true gravity $\bm{g}$ and the Newtonian gravity $\bm{g}_{_N}$ are related by
\begin{eqnarray}
	\label{Simple_interpolating_function}
	\bm{g} ~&=&~ \nu \bm{g}_{_N} \, , \text{ where} \\
	\nu ~&=&~ \frac{1}{2} + \sqrt{\frac{1}{4} + \frac{a_0}{g_{_N}}} \, .
\end{eqnarray}
In the quasi-linear formulation of MOND \citep{QUMOND}, Equation \ref{Simple_interpolating_function} is generalized to more complicated geometries by setting $\nabla \cdot \bm{g} = \nabla \cdot \left( \nu \bm{g}_{_N} \right)$. To recover the correct asymptotic limits, $\nu$ should be 1 for $g_{_N} \gg a_0$ and $\sqrt{a_0/g_{_N}}$ for $g_{_N} \ll a_0$. The transition between these limits is a free function in MOND. While several interpolating functions have been developed for this purpose \citep{Kent_1987, McGaugh_2016}, the simple interpolating function seems to work better with recent observations \citep{Iocco_Bertone_2015, Banik_2018, Chae_2018}. In particular, it provides a good fit to the relation between the radial components of $\bm{g}$ and $\bm{g}_{_N}$ in rotating galaxies \citep{McGaugh_2016, Lelli_2017}. For recent reviews of MOND, we refer the reader to \citet{Famaey_McGaugh_2012} and \citet{Milgrom_2014}.

Equation \ref{M_m-M_n} shows that dynamical masses in MOND should be lower than inferred from Newtonian dynamics. The extent to which this is true depends on what value we adopt for $r$. Since the El Gordo clusters are caught close to pericentre, we take $M$ to be their combined mass and $r$ to be the virial radius of the more massive component. Thus, we use $M_{_N} = 3.13 \times 10^{15} M_{\odot} \left( \widetilde{M}_{_N} = 15.50 \right) $ and $r = 1.65$~Mpc, with the latter coming from table~2 of \citet{Jee_2014}. Their figure~10 shows that the mass is mostly concentrated within a projected radius of 1~Mpc, so 1.65~Mpc seems like a reasonable choice for $r$ where the assumption of spherical symmetry would approximately hold. Much larger values would not be appropriate as the observations do not go out that far. With these values, we get that the weak lensing mass of El Gordo would become $M_{_M} = 1.8 \times 10^{15} M_{\odot} \left( \widetilde{M}_{_M} = 15.26 \right)$ in MOND, about half the Newtonian value.

MOND was originally developed to explain the flat rotation curves of galaxies without resorting to the presence of dark matter \citep{Milgrom_1983}. It also proved useful at explaining other phenomena such as the satellite planes of the Milky Way and Andromeda \citep{Banik_Ryan_2018, Bilek_2018}. However, MOND has faced serious problems as a model that could single-handedly explain all physical phenomena in the Universe. The most well known problem is that some undetected mass is needed to explain the velocity dispersions of galaxy clusters \citep{Sanders_1999}. Although MOND greatly reduces the discrepancy between the observed and dynamically inferred mass compared to the Newtonian case \citep{Ettori_2019}, a significant mismatch remains at small radii. The different distributions of dynamical and baryonic mass is most evident in the Bullet Cluster, where the weak lensing and X-ray peaks are offset at high significance \citep{Clowe_2006}. Therefore, the additional mass needed by MOND should be collisionless.

It is important to realise that this extra mass is evident on a much larger scale than individual galaxies, so the Bullet Cluster does not require \emph{cold} dark matter \citep{Angus_2007}. In MOND, it is of course not possible to add dark matter particles that would significantly cluster on galaxy scales. Instead, some form of hot dark matter (HDM) is required. Assuming the missing mass is not just baryons in some hard to detect form \citep{Milgrom_2008}\footnote{this would involve only a small fraction of the baryons expected from Big Bang nucleosynthesis}, a popular candidate is sterile neutrinos \citep{Angus_2009} since these do not require any significant modifications to the standard model of particle physics \citep{Merle_2017, Boyarsky_2019}. Neutrinos are also the only known massive particles which are electrically neutral and long-lived. In the MOND context, an undiscovered species of sterile neutrino would be required with a mass of $m_{\nu_s} = 11$~eV/$c^2$ because this is the mass at which thermally produced sterile neutrinos have the same average mass density as the dark matter in $\Lambda$CDM. Sterile neutrinos with $m_{\nu_s} > 10$~eV/$c^2$ have such a short free streaming length as to be consistent with the Planck results \citep[section~6.4.3 of][]{Planck_XIII_2015}. $11$~eV/$c^2$ sterile neutrinos thus provide a plausible explanation for the observed anisotropies in the CMB \citep{Angus_2009}. This marriage of MOND $+$ sterile neutrinos \citep[which we term $\nu$HDM; discussed further in section~3.1 of][]{Haslbauer_2020} can also explain the internal dynamics of 30 virialized galaxy clusters, with the implied neutrino density marginally reaching the Tremaine-Gunn limit \citep{Tremaine_Gunn_1979} at the centre once allowance is made for the brightest cluster galaxy \citep{Angus_2010}. An extra sterile neutrino could plausibly have evaded direct detection with current technology, though there are some hints for it in terrestrial experiments \citep[e.g.][]{Miniboone_2018}.

Structure formation in MOND is expected to be much more efficient than in $\Lambda$CDM \citep{Sanders_1998}. In \citet{Angus_2011} and \citet{Angus_2013}, the $\nu$HDM model was explored with cosmological $N$-body simulations that assumed a standard expansion history \citep{Skordis_2006} and applied MOND only to the density perturbations \citep{Sanders_2001, Nusser_2002}. The main conclusion of \citet{Angus_2013} was that massive galaxy clusters are overproduced in $\nu$HDM. However, this can be attributed to several factors:
\begin{enumerate}
	\item The resolution of their $\left( 256 \, h^{-1} \, \text{cMpc} \right)^3$ simulation was very low, so it could barely account for the presence of smaller, less massive structures. This makes it less likely to end up having small structures in the simulation, and more likely to have a few very massive objects.
	
	\item A small box like the one they used cannot account for the external field from distant background objects. The external field effect is a physical consequence of the non-linear gravity law in MOND \citep{Milgrom_1986}. It implies that the internal gravity of a system is weakened by a constant gravitational field from the external environment, even in the absence of tides. In this case, the presence of large background accelerations would have made the gravitational potential of the system more Newtonian, possibly suppressing the overproduction of massive structures.
	
	\item The mass function produced by the simulation was compared to the cluster mass functions of \citet{Reiprich_2002} and \citet{Rines_2008}, which were developed for galaxy clusters that mostly lie at $z \la 0.1$. According to \citet{KBC_2013}, our Galaxy lies inside an $\approx 300$~Mpc void with density contrast of $\approx -0.5$. Galaxy clusters at low $z$ would be inside this void (hereafter the KBC void). Therefore, the $z \la 0.1$ cluster mass function is very likely not representative of the Universe as a whole. This is also apparent in that El Gordo-like objects should be far more common at $z = 0$ than at $z = 1$ regardless of the cosmological model. Thus, it is surprising to not see similarly extreme objects at low $z$ (except perhaps the Bullet Cluster at $z = 0.296$). This is most likely due to the KBC void out to $z \approx 0.15$ \citep{KBC_2013}, which is also apparent in X-ray surveys of galaxy clusters \citep{Boehringer_2015, Boehringer_2020}. We point out that the \citet{Angus_2011} simulations already predicted the presence of large voids with $250 \, h^{-1}$~Mpc diameter, but attributed this to a flaw in their model. There is strong evidence that we are living inside such a supervoid \citep[i.e. the KBC void; see section~1 of][and references therein]{Haslbauer_2020}. By modelling its detailed dynamics in the $\nu$HDM framework, they showed that it can explain the Hubble tension and the curvature of the low $z$ distance-redshift relation, which are otherwise difficult to understand in $\Lambda$CDM.
\end{enumerate}


\citet{Katz_2013} explored the $\nu$HDM model using $N$-body simulations in a box whose linear dimensions were twice as large as those of \citet{Angus_2013}. The conclusion of \citet{Katz_2013} was that both the velocities and masses of massive clusters are larger at late times in MOND compared to $\Lambda$CDM. Clusters become more massive at an earlier epoch in MOND, which would help to explain the large number of massive high-$z$ clusters that have been found in the last few years \citep[e.g.][]{Foley_2011}. As with the earlier $\nu$HDM simulations, \citet{Katz_2013} also reported an overabundance of high-mass clusters. However, they again used the \citet{Reiprich_2002} observational data for comparison. This catalogue reaches clusters up to $z \approx 0.3$ but is mostly composed of lower redshift clusters, whose abundance would be significantly affected by the KBC void.

\citet{Katz_2013} found that among their simulated cluster pairs with a similar redshift, mass, and mass ratio to the Bullet Cluster, $13\%$ reach its 3000~km/s infall velocity in MOND, but only $2\%$ do so in their larger $\Lambda$CDM simulation volume (see their figure~8). Remarkably, they managed to identify a few pairs in their MOND simulation that match the redshift, collision velocity, and main cluster mass of El Gordo (see their section~5.1). Their adopted pre-merger configuration had an infall velocity of 2300~km/s and a main cluster mass of $2 \times 10^{15} M_\odot$, which as discussed above are reasonable in a MOND context. Pairs like El Gordo are quite rare at its high redshift in $\nu$HDM $-$ but they do occur. No such pairs were found in their $\Lambda$CDM simulation. Therefore, the presence of El Gordo-like objects is not as unusual in the $\nu$HDM model compared to $\Lambda$CDM.

To quantify whether the occurrence rate of El Gordo analogues in $\nu$HDM is similar to observations, we perform a simple calculation of how many El Gordo-like objects are expected in the surveyed region. For this, we assume that one pair analogous to El Gordo was discovered by \citet{Katz_2013} in their $\left( 512 \, h^{-1} \, \text{cMpc} \right)^3$ simulation volume. Given the effective survey volume calculated with Equations \ref{Delta_a} and \ref{V_eff}, we obtain that $N_{EG} = 1.16$. This assumes that the growth parameter $k$ in Equation \ref{k} ($k = 32.75$ for a fit with $\widetilde{M}_{min} = 14.16$ after requiring $\widetilde{v} > \widetilde{v}_{EG}$) is the same in MOND as in $\Lambda$CDM, which is not necessarily true. While a more accurate calculation goes beyond the scope of this contribution, we can assume that $k$ should not change by more than order unity (i.e. by $\la 1$ dex) regardless of whether we are using MOND or $\Lambda$CDM. Since $N_{EG} \appropto 1/k$ for $k \gg 1$ (Equation \ref{Delta_a}), order unity changes to $k$ should not significantly affect $N_{EG}$. It would also not make much difference if instead of 1 pair analogous to El Gordo, \citet{Katz_2013} had found e.g. 3 pairs $-$ they certainly did not find very many, but they did find $\geq 1$ (see their section~5.1). Thus, \emph{the $\nu$HDM framework yields the correct order of magnitude for the number of El Gordo analogues in the surveyed region}.

Part of the reason for this is that the MOND dynamical mass is about half the Newtonian value. As discussed in section~\ref{Lower_mass}, a lower mass helps to alleviate the tension with the simulated cluster mass function. If we repeat the redshift tomography analysis using instead $M_{_M}$ while still keeping the infall velocity and impact parameter of fiducial Model B, we obtain that the expected number of analogues rises from $N_{EG} = 7.51 \times 10^{-10}$ to $N_{EG} = 3.78 \times 10^{-6}$, reducing the tension from $6.16 \sigma$ to $4.62\sigma$. While this manages to get the tension below $5\sigma$, it rises back to $5.04\sigma$ when combined with the Bullet Cluster. Since we mentioned above that $N_{EG} \approx 1$ in MOND, the lower mass is not the main reason why it could explain the existence of El Gordo. Rather, the MOND model manages to raise the number of El Gordo analogues mainly by enhancing structure growth.

The extent to which this occurs depends on how gravitational fields from inhomogeneities couple to $g_{_{Hubble}}$, the acceleration required to maintain the time-dependent Hubble flow. We discuss this theoretical uncertainty below, and refer the reader to section~5.2.3 of \citet{Haslbauer_2020} for a more detailed discussion. The $\nu$HDM simulations of \citet{Katz_2013} applied MOND only to the gravity sourced by inhomogeneities, thus assuming no coupling to $g_{_{Hubble}}$. The possibility of a non-trivial coupling was discussed in \citet{Sanders_2001}, but has rarely been considered since. If there is such a coupling, the resulting Hubble field effect (HFE) could raise the appropriate value for the gravity in Equation \ref{Simple_interpolating_function}, thereby suppressing the MOND boost to gravity. In other words, a large background acceleration could make the gravitational potential of the system more Newtonian, which can dampen the production of massive structures. Therefore, the \citet{Katz_2013} simulation might have overestimated the number of very massive objects that we should find in a $\nu$HDM universe. The assumption of no HFE may have affected the \citet{Katz_2013} results only if:
\begin{enumerate}
	\item the gravitational field $g$ of structures that were already included in the simulation is smaller than $a_{_0}$, and
	\item $g_{_{Hubble}}$ dominates over $g$.
\end{enumerate}
The first condition is required so there is a MOND enhancement to gravity at all $-$ if not, then the HFE cannot further suppress the already non-existent MOND boost to gravity. The second condition is required for the HFE to be significant. To make further progress, we note that the $g_{_{Hubble}}$ term for a system with size $r$ is defined as:
\begin{eqnarray}
	g_{_{Hubble}} &\equiv& \frac{\ddot{a}}{a}r ~=~ -\frac{4\pi G}{3}\left(\rho_m - 2\rho_{\Lambda} \right)r \\
	&=& {H_0}^2 \left(-\frac{1}{2} a^{-3} \Omega_{m, 0} + \Omega_{\Lambda, 0}\right) r_c a \, ,
	\label{ghubble}
\end{eqnarray}
where we use the canonical $a_0 = 1.2 \times 10^{-10}$~m/s$^2$ \citep{Begeman_1991, Gentile_2011} and the same values as the Jubilee simulation for the present cosmological parameters $H_0$, $\Omega_{m, 0}$, and $\Omega_{\Lambda, 0}$. The co-moving radius $r_c = 22.70$~cMpc is the radius of a sphere enclosing the mass of El Gordo at the cosmic mean density. This is redshift-independent due to mass conservation. $r_c$ is thus a typical co-moving length scale for the problem, while $r$ is the corresponding physical scale.

To estimate $g$ sourced by inhomogeneities, we need to make a few assumptions. We take that perturbations grow $\propto a$, as occurs in $\Lambda$CDM during the matter-dominated era. This is conservative as it raises the relative importance of the HFE by suppressing $g$. With this assumption, the Newtonian gravity of the inhomogeneities is $g_{_N} \propto a^{-1}$. In order to obtain the actual value of $g_{_N}$, we need to know $g_{_N}$ at some epoch, which we take to be the epoch of recombination ($a_{_{CMB}} \approx 1/1100 = 9.1 \times 10^{-4}$). We define that $g_{_N} = g_{_{CMB}}$ at that epoch, so in general:
\begin{eqnarray}
	g_{_N} = g_{_{CMB}} \times \frac{a_{_{CMB}}}{a} \, .
	\label{g_N}
\end{eqnarray}
Following section~3.1.3 of \citet{Haslbauer_2020}, we estimate the density fluctuations at recombination to be $\delta_{CMB} \approx 10^{-4}$, implying the typical gravitational field is
\begin{eqnarray}
	g_{_{CMB}} \approx \frac{c\delta_{CMB}}{t_{_{CMB}}} \approx 21 \, a_0 \, ,
	\label{gCMB}
\end{eqnarray}
where $t_{_{CMB}} = 380$~kyr is the time of recombination. We then assume that since El Gordo is a rare object even in the MOND context, it corresponds to at least a $2\sigma$ density fluctuation. The gravitational field at recombination would thus be $g_{_{CMB}} \approx 42 \, a_0$. The MOND boost to this is negligible, but at later times the enhancement to $g_{_N}$ needs to be calculated using Equation \ref{Simple_interpolating_function}.

In the matter-dominated era, $g_{_{Hubble}} \propto a^{-2}$ (Equation \ref{ghubble}) while $g_{_N} \propto a^{-1}$. As a result, $g_{_{Hubble}}$ becomes sub-dominant to $g$ after some epoch $a = a_{_{HFE}}$. Meanwhile, the HFE plays no role if $g \ga a_0$ because the behaviour is Newtonian regardless of the HFE. Thus, the HFE can also be neglected prior to the epoch when $g = a_0$, which we define as occurring when $a = a_{_{MOND}}$. Therefore, the \citet{Katz_2013} simulation might overestimate the MOND enhancement to gravity only in the period when $a_{_{MOND}} \la a \la a_{_{HFE}}$. Using Equations \ref{ghubble} and \ref{g_N} with the interpolating function in Equation \ref{Simple_interpolating_function}, we obtain that $a_{_{HFE}} = 0.06$ while $a_{_{MOND}} = 0.08$. Hence, there is never any era during which there is a significant enhancement to Newtonian gravity which might be over-estimated by not considering a possible HFE $-$ either the HFE would be a sub-dominant correction, or the behaviour is Newtonian in any case, so the HFE could not make the growth of structure even more Newtonian. This means theoretical uncertainties regarding the HFE are very small at the 22.7~cMpc scale relevant to the formation of El Gordo analogues in the $\nu$HDM framework. As a result, their estimated frequency in the \citet{Katz_2013} simulation should be a good representation of their frequency in this framework. 

From the \citet{Katz_2013} results, it appears that the $\nu$HDM model, while still in need of improvement, could serve as a possible explanation for massive high velocity galaxy cluster collisions at high redshift. In particular, we see that \citet{Angus_2013} were not justified in their criticism that MOND does not get the correct cluster mass function at the high mass end. Leaving aside other issues like resolution, if massive clusters were over-produced, then massive cluster pairs would certainly be over-produced. However, our calculations show that \citet{Katz_2013} obtained approximately the correct number of El Gordo analogues in their simulation volume. This conclusion should remain valid despite uncertainty in how the Hubble acceleration couples to that sourced by inhomogeneities in a MOND universe. Therefore, it is important to revisit the $\nu$HDM model.

\section{Conclusions}
\label{Conclusions}

We studied the probability of observing an El Gordo-like collision between two massive high-$z$ galaxy clusters by searching for progenitor galaxy cluster pairs that have turned around from the cosmic expansion in the $\left( 6 \, h^{-1} \, \text{cGpc} \right)^3$ Jubilee simulation box at $z = 1$. We used two different statistical analyses based on the number of analogue pairs to the El Gordo progenitor clusters with appropriate total mass, mass ratio, and pre-merger infall velocity. The initial conditions were obtained from the hydrodynamical simulations of \citet{Zhang_2015}, from which we adopt their nominal Model B. Our main result is that the discovery of an El Gordo-like collision within the surveyed region excludes the $\Lambda$CDM model at $6.16\sigma$ using the method that gives less tension (lightcone tomography). We also used the analysis of \citet{Kraljic_2015} to show that the Bullet Cluster is in $2.78\sigma$ tension with $\Lambda$CDM. In this framework, the combination of these observations is a highly unlikely $6.43\sigma$ event.


We considered whether El Gordo could have a lower mass (Section~\ref{Lower_mass}). However, this only reduces the tension to $5.14\sigma$ ($5.50\sigma$ when combined with the Bullet Cluster). In reality, the tension should be even higher as it does not account for the following:
\begin{enumerate}
	\item the poor fit to the observed twin-tailed morphology of El Gordo,
	\item the discrepancy between the assumed mass and the weak lensing observations \citep{Jee_2014}, and
	\item the small impact parameter required in this scenario, which may be unlikely on geometrical grounds.
\end{enumerate}
We also consider the possibility that El Gordo and the Bullet Cluster are extremely unique objects such that no other problematic objects like these will be found in the remaining observable sky. Increasing the surveyed area to the full sky reduces the combined tension of El Gordo and the Bullet Cluster to $5.66\sigma$, which does not contribute that much to solve the problem. Besides, the assumption is not very realistic $-$ full sky surveys like Planck have already found interacting clusters whose properties are unlikely to arise in a $\Lambda$CDM context (Section~\ref{Combined_with_BC}). Further work is required to quantify the extent to which the cases discussed there and in Section \ref{Introduction} are problematic for $\Lambda$CDM.

Explaining the Bullet Cluster, El Gordo, and a growing number of fast, massive, high-redshift galaxy clusters can be very difficult within the $\Lambda$CDM cosmology. An alternative model uses MOND gravity supplemented by 11~eV/$c^2$ sterile neutrinos to explain various phenomena including the CMB anisotropies, cluster-scale problems for purely baryonic MOND, and the baryon-weak lensing offset in the Bullet Cluster \citep[Section~\ref{MOND}; see also section~3.1 of][]{Haslbauer_2020}. This $\nu$HDM model was previously explored in \citet{Katz_2013}, where a handful of objects analogous to El Gordo were found in their simulation volume of $\left( 512 \, h^{-1} \, \text{cMpc} \right)^3$ (see their section~5.1). This is very similar to the effective volume of the survey which discovered El Gordo. We estimate that the \citet{Katz_2013} simulation implies 1.16 El Gordo analogues in the surveyed volume, implying good agreement with observations. Although this result could be off by a factor of a few, it is clear that the $\nu$HDM model produces approximately the correct abundance of massive cluster pairs similar to the El Gordo progenitors. Thus, we argue against the conclusion of \citet{Angus_2013} that MOND overproduces massive galaxy clusters. Section~\ref{MOND} discusses possible reasons for their erroneous conclusion.

We conclude that the El Gordo galaxy cluster collision rules out the $\Lambda$CDM cosmology at high significance, but can likely be explained in a MOND cosmology supplemented by 11~eV/$c^2$ sterile neutrinos. While it is difficult to dispute the evidence for dark matter in galaxy clusters like the Bullet, it is also difficult to explain its properties and those of El Gordo without modifying gravity on large scales $-$ or at low accelerations. Given also the galaxy-scale challenges for $\Lambda$CDM \citep[e.g.][]{Kroupa_2012, Kroupa_2015} and the many successes of MOND on this scale \citep[e.g.][]{Famaey_McGaugh_2012, Li_2018}, the most likely scenario is that at present we have understood neither the full matter-energy content of the Universe nor the law of gravity that governs it.

\section*{Data availability}

The data underlying this article are available in the article.

\section*{Acknowledgements}

EA is supported by a stipend from the Stellar Populations and Dynamics Research Group at the University of Bonn. IB is supported by an Alexander von Humboldt Foundation postdoctoral research fellowship. The authors would like to thank the Jubilee team for providing the data needed to carry out this project. They also thank Moti Milgrom for helpful discussions on MOND in galaxy clusters. The authors are grateful to the referee for her/his constructive comments.

\bibliographystyle{mnras}
\bibliography{ElGordo_bbl}

\begin{appendix}

\section{Mass functions for El Gordo-like cluster pairs}
\label{Appendix}

The coefficients of our quadratic fit to the cumulative mass function (Equation \ref{Mass_function_fit}) are given in Table~\ref{tab:coeffB} based on the El Gordo parameters in Model B of \citet{Zhang_2015}. The corresponding results for their Model A are given in Table~\ref{tab:coeffA}. As described in the text, both fits are done only for pairs with $\widetilde{M} > 14.16$ to avoid edge effects at low masses. Since the Jubilee simulation lacks pairs as massive as El Gordo, our statistical analysis is based on extrapolating these fits to the El Gordo mass.

\begin{table}
	\centering
		\begin{tabular}{|c|c|c|c|}
			\hline
			\multirow{2}{*}{\begin{tabular}[c]{@{}c@{}}Mass\\ function\end{tabular}} & \multicolumn{3}{c|}{Redshift} \\ \cline{2-4} 
			& $0$        & $0.509$    & $1$       \\ \hline
			$c_0$                                                                    & $-369.69$  & $-544.03$  & $-786.14$ \\ 			$c_1$                                                                    & $~~~~55.06$    & $~~~~80.05$    & $~~114.75$  \\
			$c_2$                                                                    & $~-2.02$    & $~-2.92$    & $~-4.16$   \\ \hline
		\end{tabular}
	\caption{Coefficients of our quadratic fit to the cumulative mass distribution (Equation \ref{Mass_function_fit}) for pairs which satisfy the turnaround, mass ratio, and $\widetilde{v}$ conditions appropriate to Model B of \citet{Zhang_2015}. This is the nominal case considered in our work.}
	\label{tab:coeffB}
\end{table}

\begin{table}
	\centering
		\begin{tabular}{|c|c|c|c|}
			\hline
			\multirow{2}{*}{\begin{tabular}[c]{@{}c@{}}Mass\\ function\end{tabular}} & \multicolumn{3}{c|}{Redshift} \\ \cline{2-4} 
			& $0$        & $0.509$    & $1$       \\ \hline
			$c_0$                                                                    & $-160.13$  & $-203.90$  & $-347.02$ \\ 			$c_1$                                                                    & $~~~~26.45$    & $~~~~33.46$    & $~~~~54.41$  \\
			$c_2$                                                                    & $~-1.05$    & $~-1.33$    & $~-2.10$   \\ \hline
		\end{tabular}

	\caption{Similar to Table~\ref{tab:coeffB}, but for Model A of \citet{Zhang_2015}.}
	\label{tab:coeffA}
\end{table}

\end{appendix}

\bsp
\label{lastpage}
\end{document}